\RequirePackage[T1]{fontenc}
\documentclass[copyright,creativecommons]{eptcs}

\providecommand{\firstpage}{1}

\usepackage[dvipsnames]{xcolor}
\definecolor{burntumber}{rgb}{21, 96, 189}
\usepackage{stmaryrd,paralist}
\usepackage{xspace,url,hyperref}
\usepackage{cite,enumitem,paralist}
\usepackage{mathtools}
\usepackage{amsfonts,amssymb,amsmath}
\usepackage{threeparttable,url,breakurl}
\usepackage{listings}
\usepackage[english]{babel} 
\usepackage{tikz}
\usepackage{mathpartir}
\usepackage{marginnote}
\usetikzlibrary{cd}

\newcommand{\comment}[1]{}

\begin{document}
\title{A Beluga Formalization of the\\ Harmony Lemma in the $\pi$-Calculus}
\author{Gabriele Cecilia
\institute{Dipartimento di Matematica,\\
  Universit\`{a} degli Studi di Milano, Italy}
  \and
Alberto Momigliano
 \institute{Dipartimento di Informatica,\\
  Universit\`{a} degli Studi di Milano, Italy}}
 \def\authorrunning{Cecilia \& Momigliano}
 \def\titlerunning{Formalizing the Harmony Lemma in Beluga}
\maketitle
\begin{abstract}
  The ``Harmony Lemma'', as formulated by Sangiorgi \& Walker,
  establishes the equivalence between the labelled transition
  semantics and the reduction semantics in the $\pi$-calculus. Despite
  being a widely known and accepted result for the standard
  $\pi$-calculus, this assertion has never been rigorously proven,
  formally or informally. Hence, its validity may not be immediately
  apparent when considering extensions of the
  $\pi$-calculus. Contributing to the second challenge of the
  Concurrent Calculi Formalization Benchmark --- a set of challenges
  tackling the main issues related to the mechanization of concurrent
  systems --- we present a
  formalization of this result for the fragment of the $\pi$-calculus
  examined in the Benchmark. Our formalization is implemented in
  Beluga and draws inspiration from the HOAS formalization of the LTS
  semantics popularized by Honsell et al. In passing, we
  introduce a couple of useful encoding techniques for handling
  telescopes and lexicographic induction.
\end{abstract}
\long\def\ednote#1{\footnote{[{\it #1\/}]}\message{ednote!}}
\newenvironment{metanote}{\begin{quote}\message{note!}[\begingroup\it}%
                         {\endgroup]\end{quote}}
\long\def\ignore#1{}

\newtheorem{theorem}{Theorem}
\newtheorem{lemma}{Lemma}[theorem]
\newtheorem{corollary}{Corollary}[theorem]
\newtheorem{innersubstlemma}{Lemma}
\newcounter{substlemmacounter}
\setcounter{substlemmacounter}{0}
\newenvironment{substlemma}
{\stepcounter{substlemmacounter}
   \renewcommand\theinnersubstlemma{S\arabic{substlemmacounter}}
   \innersubstlemma}
{\endinnersubstlemma}

\newcommand*{\myfont}{\fontfamily{cmss}\selectfont}
\newcommand*{\myfonttwo}{\fontfamily{lmtt}\selectfont}
\newcommand*{\ovA}[1]{%
  $\m@th\overline{\mbox{#1}\raisebox{2.3mm}{}}$%
}

\definecolor{belugapurple}{RGB}{154,0,214}
\definecolor{belugared}{RGB}{225, 0, 0}
\definecolor{belugagreen}{RGB}{31,147,15}
\definecolor{belugablue}{RGB}{0,0,238}
\definecolor{belugapink}{RGB}{237,103,166}
\newcommand\delimone{{\color{belugapurple}\myfonttwo\bfseries LF}\color{belugagreen}\aftergroup:}
\newcommand\delimtwo{{\color{belugapurple}\myfonttwo\bfseries inductive}\color{belugagreen}\aftergroup:}
\newcommand\delimthree{{\color{belugapurple}\myfonttwo\bfseries schema}\color{belugagreen}\aftergroup=}
\newcommand\delimfour{{\color{belugapurple}\myfonttwo\bfseries rec}\color{belugablue}\aftergroup:}
\newcommand\delimfive{{/}\slshape\aftergroup/}
\newcommand\delimaux{{\color{belugapink}.}}
\newcommand\delimsix{\color{belugapink}--\aftergroup\delimaux}
\newcommand\delimseven{{\color{belugapurple}\myfonttwo\bfseries and} {\color{belugagreen}late\_bstep}\aftergroup:}
\newcommand\delimeight{{\color{belugapurple}\myfonttwo\bfseries and} {\color{belugagreen}early\_bstep}\aftergroup:}
\newcommand\delimnine{{\color{belugapurple}\myfonttwo\bfseries and} {\color{belugagreen}bstep}\aftergroup:}

\lstdefinelanguage{Beluga}
{
  morekeywords=[1]{mlam,fn,case,of,total,in,type, impossible, let, and,ctype},
  keywordstyle=[1]\color{belugapurple}\myfonttwo\bfseries,
  morecomment=[l]{\%},
  morecomment=[s]{\%\{}{\}\%},
  commentstyle=\color{belugared},
  sensitive=true,
}

\lstdefinestyle{belugastyle}{
 language=Beluga,
    basicstyle=\ttfamily\small,
    columns=flexible,
    keepspaces=true,
    showstringspaces=false,
    breaklines=true,
    breakatwhitespace=true,
    rangeprefix=\%\%\ ,
    rangesuffix=\ \%\%,
    includerangemarker=false,
    moredelim=**[is][\delimone]{LF}{:},
    moredelim=**[is][\delimtwo]{inductive}{:},
    moredelim=**[is][\delimthree]{schema}{=},
    moredelim=**[is][\delimseven]{and\ late\_bstep}{:},
    moredelim=**[is][\delimeight]{and\ early\_bstep}{:},
    moredelim=**[is][\delimnine]{and\ bstep}{:},
    moredelim=**[is][\delimfour]{rec}{:},
    moredelim=**[is][\delimfive]{/}{/},
    moredelim=**[is][\delimsix]{--}{.},
    literate={→}{{$\rightarrow$}}{1}
                  {⊢}{{$\vdash$}}{1}
                  {⇒}{{$\Rightarrow$}}{1}
}

\lstdefinestyle{belugatwo}{
 language=Beluga,
    basicstyle=\ttfamily\small,
    columns=flexible,
    keepspaces=true,
    showstringspaces=false,
    breaklines=true,
    breakatwhitespace=true,
    rangeprefix=\%\%\ ,
    rangesuffix=\ \%\%,
    includerangemarker=false,
    moredelim=**[is][\delimone]{LF}{:},
    moredelim=**[is][\delimtwo]{inductive}{:},
    moredelim=**[is][\delimthree]{schema}{=},
    moredelim=**[is][\delimseven]{and\ late\_bstep}{:},
    moredelim=**[is][\delimeight]{and\ early\_bstep}{:},
    moredelim=**[is][\delimnine]{and\ bstep}{:},
    moredelim=**[is][\delimfour]{rec}{:},
    moredelim=**[is][\delimfive]{/}{/},
    moredelim=**[is][\delimsix]{--}{.},
    frame=tblr,
    captionpos=b,
    literate={→}{{$\rightarrow$}}{1}
                  {⊢}{{$\vdash$}}{1}
                  {⇒}{{$\Rightarrow$}}{1}
}

\lstset{style=belugastyle}

\newcommand{\ExternalLink}{%
    \tikz[x=1.2ex, y=1.2ex, baseline=-0.05ex]{%
        \begin{scope}[x=1ex, y=1ex]
            \clip (-0.1,-0.1) 
                --++ (-0, 1.2) 
                --++ (0.6, 0) 
                --++ (0, -0.6) 
                --++ (0.6, 0) 
                --++ (0, -1);
            \path[draw, 
                line width = 0.5, 
                rounded corners=0.5] 
                (0,0) rectangle (1,1);
        \end{scope}
        \path[draw, line width = 0.5] (0.5, 0.5) 
            -- (1, 1);
        \path[draw, line width = 0.5] (0.6, 1) 
            -- (1, 1) -- (1, 0.6);
        }
    }
    
\setlength{\marginparsep}{2pt}

\section{Introduction}\label{sec:intro}
At page 51 of their ``bible'' on the
$\pi$-calculus~\cite{DBLP:books/daglib/0004377}, Sangiorgi \& Walker
state the \emph{Harmony Lemma}, regarding the relationship between the
reduction semantics and the transitional one (LTS). The sketch of the proof
starts as follows:
\begin{quote}
  \emph{Rather than giving the whole (long) proof, we explain the strategy and invite the reader to check some of the details [\dots]}
\end{quote}
While this informal style of proof, akin to the infamous “proof on a
napkin” championed by de Millo and colleagues\footnote{``Social
  Processes and Proofs of Theorems and Programs’', CACM 22-5, 1979.},
may be suitable for a (long) textbook, it might not be applicable to
emerging calculi with more unconventional operational semantics. Although
the theorem is undisputed within the well-established framework of the
$\pi$-calculus, this assurance may not extend to these developing
calculi. In such instances, a more rigorous approach, potentially in
the form of a machine-checked proof, is advisable.

These considerations are of course not novel: they have been
prominently argued for in the POPL\-Mark challenge~\cite{poplmark} and
subsequent follow-ups~\cite{FeltyMP18,POPLMarkReloaded}.
The recent Concurrent Calculi Formalization Benchmark~\cite{ConcBench}
(CCFB in brief) introduces a new collection of benchmarks addressing
 challenges encountered during the mechanization of 
models of concurrent and distributed programming languages, with an
emphasis on process calculi. As with POPLMark, the idea is to explore
the state of the art in the formalization in this subarea, finding
the best practices to address their typical issues and improving the
tools for their mechanization.

CCFB considers {in isolation} three aspects that may be problematic
when mechanizing concurrency theory: \emph{linearity}, \emph{scope
  extrusion}, and \emph{coinductive reasoning}. Scope extrusion is, of
course,  the method by which a process can transfer
restricted names to another process, as long as the restriction can be
safely expanded to include the receiving process.
 This phenomenon has been captured in two
different, yet equivalent ways of  formulating the operational semantics of the
$\pi$-calculus:
  \begin{enumerate}
  \item a reduction system, which avoids explicit reasoning about scope
  extrusion by using structural congruence;
\item a labelled transition system, which introduces a new kind of action to  handle extrusion directly: in doing so, it  breaks
  shared conventions such as $\alpha$-equivalence.\footnote{There are also intermediate
  approaches that save $\alpha$-equivalence, such as Parrow's LTS with structural congruence~\cite{DBLP:books/el/01/Parrow01} or Milner's notion of abstraction and concretion as formalized for example in~\cite{Bengtson2009}.}
\end{enumerate}
%
%
The second challenge in the Concurrent Calculi Formalization Benchmark
(CCFB.2) consists in mechanizing these two operational semantics and
relating them via the aforementioned Harmony Lemma.

Obviously, we are not the first to address the mechanization of the
$\pi$-calculus (although we seem to be the first to tackle the Harmony
result): given the challenges that it poses (various kind of binders
with somewhat unusual properties compared to the $\lambda$-calculus),
there is a long tradition starting with~\cite{Melham1994} and mostly
developed with encodings based on first-order syntax such as de Brujin
indexes --- see~\cite{ConcBench} for a short review of the literature
w.r.t.\ scope extrusion. As often remarked, concrete encodings will
get you there, but not effortlessly: an estimation of $75$\% of the
development being devoted to the infrastructure of names handling is
not uncommon~\cite{Hirschkoff97}:
\begin{quote}
  ``Technical work, however, still represents the biggest part of our
  implementation, mainly due to the managing of De Bruijn indexes [\dots] 
  Of our 800 proved lemmas, about 600 are concerned with
  operators on free names.''
\end{quote}

It is not surprising that specifications based on higher-order
abstract syntax (HOAS) soon emerged, first only as animations,
see~\cite{MillerPI} in $\lambda$Prolog and~\cite{HonsellLMP98} in
LF\@. Moving to meta-reasoning, we can roughly distinguish two main approaches:
\begin{enumerate}
\item ``squeezing'' HOAS into a general proof assistant: there is a
  plethora of approaches, but  w.r.t\ the $\pi$-calculus this has been 
  investigated by Despeyroux~\cite{Despeyroux00} and then
  systematically by Honsell and his colleagues, starting
  with~\cite{DBLP:journals/tcs/HonsellMS01} and then addressing other
  calculi;
  
\item the Pfenning-Miller ``two-level approach'' of separating the
  specification from the reasoning logic, whose culmination, as far as
  the $\pi$-calculus is concerned, is the most elegant version
  presented in~\cite{DBLP:journals/tocl/TiuM10} and later implemented
  in Abella.
\end{enumerate}

We fall in the second camp and we offer a
Beluga~\cite{DBLP:conf/cade/PientkaD10} mechanization of CCFB.2
together with a detailed informal proof, filling all the gaps left by
the quoted sketch.  Along the way, we introduce (or simply rediscover)
a couple of Beluga tricks to encode \emph{telescopes} (i.e.\ n-ary
sequences of binders) and to simulate lexicographic induction.
We also prove another folk
result, namely the equivalence between the early and late LTS, as well
as what is sometimes called ``internal
adequacy''~\cite{DBLP:journals/tcs/HonsellMS01}, that is the
equivalence between the LTS encoding from the Honsell paper with the
one in~\cite{DBLP:journals/tocl/TiuM10}.

Informal and formal proofs in all their glory can be found
here~\cite{GBThesis}. In the text, the statements of informal lemmas and theorems are hyperlinked
to their formalization in the repository.
For reasons of space, we will assume familiarity with the basic
notions of the $\pi$-calculus as in~\cite{DBLP:books/el/01/Parrow01},
as well as a working knowledge of Beluga, both of its syntax and more
importantly of its approach to proof checking.


\section{The $\pi$-Calculus and its Operational Semantics}\label{sec:picalc}
In this section, we quickly recall the main notions involved, so as to make
the paper self-contained. For more details
see~\cite{DBLP:books/daglib/0004377}.
\subsection{Syntax}
We assume the existence of a countably infinite set of \emph{names},
ranged over by $x,y,\dots$ We make no other assumption about names,
since the syntax of \emph{processes} in CCFB.2 does not consider
(mis)match. In fact, to concentrate in isolation on scope extrusion,
sums and replications are ignored as well:
\begin{equation*}
  P, Q \ \vcentcolon \vcentcolon = \ \textbf{0} \ \mid \ x(y).P \ \mid \ \bar{x}y.P \ \mid \ (P \mid Q) \ \mid \ (\nu x) P
\end{equation*} 
The input prefix $x(y).P$ and the restriction $(\nu y) P$ both bind
the name $y$ in $P$. 
Any other
occurrence of names in a process is free. The sets of free and bound
names occurring in a process ({\myfont fn($P$)} and {\myfont bn($P$)}
respectively) are defined as usual.

In the mathematical presentation of the operational semantics, we
adopt the following slightly weaker variable
convention\footnote{Variable conventions are used in a rather loose
  way in the literature, e.g.\  Parrow
  and Sangiorgi \& Walker
  adopt the
  same convention, but end up with  different provisos  in the operational semantics rules.}:
\begin{inparaenum}[1)]
	\item given a process, it is \emph{possible} to $\alpha$-rename the bound occurrences of variables within it;
	\item the bound names of any processes or actions under consideration \emph{can} be chosen different from the names occurring free in any other entities under consideration.
\end{inparaenum}

\subsection{Reduction Semantics}
We define \emph{structural congruence}  ($\equiv$) and 
\emph{reduction}  ($\rightarrow$) as the smallest binary
relations over processes, respectively satisfying the axioms in
Fig.~\ref{fig:congred}.  The notation $Q \{ y/z \}$ represents
capture-avoiding substitution of $y$ for $z$ in the process $Q$. Note
that we have chosen to present congruence as the compatible refinement
of the six basic axioms, rather than using process \emph{contexts} as
in~\cite{DBLP:books/daglib/0004377}, since the latter tend to be
problematic w.r.t.\ a HOAS formalization.

\begin{figure}[th]
  \begin{small}
\centering
\fbox{\vbox {\advance \hsize by -2\fboxsep \advance \hsize by -2\fboxrule \linewidth\hsize
\begin{mathpar}
  \inferrule[$\mkern 40mu$ Par-Assoc] { } {P \mid (Q \mid R) \, \equiv \, (P \mid Q) \mid R}
  \mkern 92mu \inferrule[Par-Unit] { } {P \mid \textbf{0} \, \equiv \, P}
  \mkern 32mu \and \mkern 8mu \inferrule[$\mkern 2mu$ Par-Comm] { } {P \mid Q \, \equiv \, Q \mid P} \mkern 42mu \\
  \inferrule[] { } { }
  \mkern 35mu \inferrule[Sc-Ext-Zero] { } {\ (\nu x)\, \textbf{0} \, \equiv \, \textbf{0} \ \ }
  \mkern 76mu \inferrule[$\mkern 38mu$ Sc-Ext-Par] {x \notin$ {\myfont fn}$(Q)} {(\nu x) P \mid Q \, \equiv \, (\nu x) (P \mid Q)}
  \mkern 29mu \inferrule[\qquad \ \ Sc-Ext-Res] { } {(\nu x) (\nu y) P \, \equiv \, (\nu y) (\nu x) P} \\
  \vspace{-0.5\baselineskip} \hbox to 16cm{\leaders\hbox to 10pt{\hss - \hss}\hfil} \\
  \inferrule[$\mkern 38mu$ C-In] {P \equiv Q} {x(y).P \, \equiv \, x(y).Q}
  \and \inferrule[$\mkern 17mu$ C-Out] {P \equiv Q} {\bar{x}y.P \, \equiv \, \bar{x}y.Q}
  \and \inferrule[$\mkern 24mu$ C-Par] {P \equiv P'} {P \mid Q \, \equiv \, P' \mid Q}
  \and \inferrule[$\mkern 31mu$ C-Res] {P \equiv Q} {(\nu x) P \, \equiv \, (\nu x) Q} \\ 
  \vspace{-0.5\baselineskip} \hbox to 16cm{\leaders\hbox to 10pt{\hss - \hss}\hfil} \\
  \inferrule[] { } { }
  \mkern 45mu \inferrule[C-Ref] { } {P \equiv P}
  \and \mkern 74mu \inferrule[C-Sym] {P \equiv Q} {Q \equiv P} 
  \and \mkern 31mu \inferrule[$\mkern 25mu$ C-Trans] {P \equiv Q \\ Q \equiv R} {P \equiv R} \mkern 10mu \\
  \hbox to 16cm{\leaders\hbox to 10pt{---}\hfil} \\
  \inferrule[$\mkern 75mu$ R-Com] { } {\bar{x}y.P \mid x(z).Q \ \rightarrow \ P \mid Q \{ y/z \}} 
  \and \inferrule[$\mkern 24mu$ R-Par] {P \rightarrow Q} {P \mid R \, \rightarrow \, Q \mid R} \mkern 55mu \\
  \inferrule[] { } { } \mkern 56mu
  \inferrule[$\mkern 33mu$ R-Res] {P \rightarrow Q} {(\nu x)P \, \rightarrow \, (\nu x)Q}
  \mkern 92mu \inferrule[$\mkern 78mu$ R-Struct] {\,P \equiv P' \\ \,\,P' \rightarrow Q' \\ Q' \equiv Q} {P \rightarrow Q}
\end{mathpar}
}}
\end{small}
\vspace{-5mm}
\caption{Congruence and reduction rules.}
\label{fig:congred}
\end{figure}

\subsection{Labelled Transition System Semantics}
The syntax of \emph{actions} is the following:
\begin{equation*}
  \alpha \, \vcentcolon = \ x(y) \ \mid \ \bar{x}y \ \mid \ \bar{x}(y) \ \mid \ \tau
\end{equation*}
In the input action $x(y)$ and in the bound output action
$\bar{x}(y)$, the name $x$ is free and $y$ is bound; in the free
output action $\bar{x}y$, both $x$ and $y$ are free. 
The sets of
free names, bound names and names occurring in an action ({\myfont
bn($\alpha$)}, {\myfont fn($\alpha$)} and {\myfont n($\alpha$)}
respectively) are defined
accordingly.
The \emph{transition} relation $\cdot \xrightarrow{\cdot}\cdot$ is the
smallest relation which satisfies the rules in Fig.~\ref{fig:lts}.
\begin{figure}[ht]
\centering
\begin{small}
\fbox{\vbox {\advance \hsize by -2\fboxsep \advance \hsize by -2\fboxrule \linewidth\hsize
\begin{mathpar}
 \inferrule[$\mkern 30mu$ S-In] { } {x(z).P \, \xrightarrow{x(z)} \, P}
  \and \mkern 75mu \inferrule[$\mkern 9mu$ S-Out] { } {\bar{x}y.P \, \xrightarrow{\bar{x}y} \, P} \mkern 16mu \\
  \inferrule[$\mkern 77mu$ S-Par-L] {P \xrightarrow{\alpha} P' \\ ${\myfont bn}$(\alpha)$ $\cap$ {\myfont fn}$(Q) = \emptyset} {P \mid Q \ \xrightarrow{\alpha} \ P' \mid Q}
  \mkern 68mu \inferrule[$\mkern 77mu$ S-Par-R] {Q \xrightarrow{\alpha} Q' \\ ${\myfont bn}$(\alpha)$ $\cap$ {\myfont fn}$(P) = \emptyset} {P \mid Q \ \xrightarrow{\alpha} \ P \mid Q'} \\
  \inferrule[$\mkern 44mu$ S-Com-L] {P \xrightarrow{\bar{x}y} P' \\ Q \xrightarrow{x(z)} Q'} {P \mid Q \, \xrightarrow{\tau} \, P' \mid Q'\{  y/z \}} \mkern 13mu
  \and \inferrule[$\mkern 44mu$ S-Com-R] {P \xrightarrow{x(z)} P' \\ Q \xrightarrow{\bar{x}y} Q'} {P \mid Q \, \xrightarrow{\tau} \, P' \{ y/z \} \mid Q'} \\
  \inferrule[$\mkern 50mu$ S-Res] {P \xrightarrow{\alpha} P' \\ z \notin$ {\myfont n}$(\alpha)} {(\nu z) P \, \xrightarrow{\alpha} \, (\nu z) P'}
  \and \mkern 22mu \mkern 8mu \inferrule[$\mkern 32mu$ S-Open] {P \xrightarrow{\bar{x}z} P' \\ z \neq x} {(\nu z)P \, \xrightarrow{\bar{x}(z)} \, P'} \mkern 12mu \\
  \inferrule[$\mkern 38mu$ S-Close-L] {P \xrightarrow{\bar{x}(z)} P' \\ Q \xrightarrow{x(z)} Q'} {P \mid Q \ \xrightarrow{\tau} \ (\nu z) (P' \mid Q')}
  \and \inferrule[$\mkern 38mu$ S-Close-R] {P \xrightarrow{x(z)} P' \\ Q \xrightarrow{\bar{x}(z)} Q'} {P \mid Q \ \xrightarrow{\tau} \ (\nu z) (P' \mid Q')}
\end{mathpar}
}}
\end{small}
\vspace{-5mm}
\caption{Transition rules.}
\label{fig:lts}
\end{figure}

Unlike the reduction semantics, the transitional semantics directly addresses
scope extrusion via the two {\small \textsc{S-Close}} rules in
interaction with {\small \textsc{S-Open}}: recall how the former rules
are \emph{not} closed under $\alpha$-conversion, since the bound name
$z$ must occur free in the other premise. 

The LTS  introduced here is the \emph{late} semantics, as opposed
to the \emph{early} one adopted by the Benchmark. However, as remarked
in~\cite{DBLP:books/el/01/Parrow01}, ``it is a matter of taste which
semantics to adopt''. We indeed prove this equivalence in Appendix~\ref{app:equiv}.

\subsection{The Harmony Lemma}
\label{ssec:harmony}
In~\cite{DBLP:books/daglib/0004377}, the Harmony Lemma reads as:
\begin{enumerate}
	\item[i.] $P \equiv \xrightarrow{\alpha} Q$ implies $P \xrightarrow{\alpha} \equiv Q$.
	\item[ii.]  $P \xrightarrow{\tau}\equiv Q$  iff $P \rightarrow Q$.
\end{enumerate}
The juxtaposition of symbols denotes relational composition (e.g.\ $P \equiv \xrightarrow{\alpha} Q$ denotes $P \equiv R$ and $R \xrightarrow{\alpha} Q$ for some $R$). The first assertion is a direct consequence of Lemma~\ref{lm26}, as detailed at page~\pageref{lm26},
which is instrumental to prove the right-to-left direction of the
equivalence result. The latter breaks down into the following
theorems:
\begin{enumerate}
\item Every transition through a $\tau$ action corresponds to a reduction;
\item Given a reduction of $P$ to $Q$, $P$ is able to make a
  $\tau$-transition to some $Q'$ congruent to $Q$.
\end{enumerate}
In the interest of setting the stage for anybody who wishes to give a
solution to CCFB.2, we start by stating a few technical lemmas about
substitutions that are used in both
directions of the Harmony Lemma, while being often left unsaid. 

\begin{substlemma}
  \label{le:s1}
$Q \{x/x\} = Q$.
\end{substlemma}
\begin{substlemma}
  \label{le:s2}
If $x \notin $ {\myfont fn}($Q$), then $Q \{y/x\} = Q$.
\end{substlemma}
These two lemmas are proved by induction on the structure of the process $Q$. A consequence of the latter is the following: if $x \notin $ {\myfont fn}($Q$), then $P \{y/x\} \mid Q = (P \mid Q) \{y/x\}$.

Finally, we state a stability result
for structural congruence under substitutions, only used in the
second direction of Harmony:
\begin{substlemma}
    \label{le:s3}
If $P \equiv Q$, then $P\{y/x\} \equiv Q\{y/x\}$.
\end{substlemma}
This lemma is proved by induction on the structure of the given derivation.

\subsubsection{Theorem~\ref{thm1}: $\tau$-Transition Implies Reduction}
The proof of the first direction relies on three key lemmas which describe rewriting (up to structural congruence) of processes involved 
 in input and  output transitions.
\addtocounter{theorem}{1}
\begin{lemma} \label{lm11}
  \sloppypar
  \reversemarginpar\marginnote{\href{https://github.com/GabrieleCecilia/concurrent-benchmark-solution/blob/main/code/2_input_rewriting.bel}{\ExternalLink}}
  If $Q \xrightarrow{x(y)} Q'$ then there exist a finite (possibly
  empty) set of names $w_1,\ldots, w_n$ (with $x, y \neq w_i$
  $\forall i=1,\ldots, n$) and two processes $R, S$ 
  such that 
  $Q \equiv (\nu w_1)\ldots (\nu w_n) (x(y).R \mid S)$ and
  \mbox{$Q' \equiv (\nu w_1)\ldots (\nu w_n) (R \mid S)$}.
\end{lemma} 
\begin{lemma} \label{lm12}
\sloppypar
  \reversemarginpar\marginnote{\href{https://github.com/GabrieleCecilia/concurrent-benchmark-solution/blob/main/code/3_free_output_rewriting.bel}{\ExternalLink}}
  If $Q \xrightarrow{\bar{x}y} Q'$ then there exist a finite (possibly
  empty) set of names $w_1,\ldots, w_n$ (with $x, y \neq w_i$
  $\forall i=1,\ldots, n$) and two processes $R, S$ such that 
  $Q \equiv (\nu w_1)\ldots (\nu w_n) (\bar{x}y.R \mid S)$  and
  \mbox{$Q' \equiv (\nu w_1)\ldots (\nu w_n) (R \mid S)$}.
\end{lemma}
\begin{lemma} \label{lm13}
\sloppypar
\reversemarginpar\marginnote{\href{https://github.com/GabrieleCecilia/concurrent-benchmark-solution/blob/main/code/4_bound_output_rewriting.bel}{\ExternalLink}}
If $Q \xrightarrow{\bar{x}(z)} Q'$ then there exist a finite (possibly empty) set of names $w_1,\ldots, w_n$ (with \mbox{$x \notin \{z,w_1,\ldots,w_n\}$}) and two processes $R, S$ such that 
$Q \equiv (\nu z)(\nu w_1)\ldots (\nu w_n) (\bar{x}z.R \mid S)$  and \mbox{$Q' \equiv (\nu w_1)\ldots (\nu w_n) (R \mid S)$}.
\end{lemma}

These three lemmas are proved by induction over the structure of the
given transition. We observe that the presence of a sequence
of 
binders is not an issue in the informal presentation; on the other
hand, from the mechanization point of view, these sequences are
challenging to encode in a framework where the meta-level binder is unary.

\addtocounter{theorem}{-1}

\begin{theorem} \label{thm1}
\marginnote{\href{https://github.com/GabrieleCecilia/concurrent-benchmark-solution/blob/main/code/5_theorem1.bel}{\ExternalLink}}
$P \xrightarrow{\tau} Q$ implies $P \rightarrow Q$.
\end{theorem}

The theorem is proved by induction on the structure of the given
transition. 
If the latter  consists of  an explicit interaction of processes in a
parallel
composition, 
we apply the aforementioned lemmas to rewrite processes involved in
specific transitions up to congruence; we then 
construct  the desired reduction through a chain of congruence and
reduction rules.

\begin{corollary}
  $P \xrightarrow{\tau}\equiv Q$  entails $P \rightarrow Q$.
\end{corollary}

\subsubsection{Theorem~\ref{thm2}: Reduction Implies $\tau$-Transition}

The other direction 
starts
with five technical lemmas regarding free and bound names in specific
transitions. They are instrumental, together with the variable
convention, to the firing of the appropriate transitions.

\addtocounter{theorem}{1}
\begin{lemma} \label{lm21}
If $P \xrightarrow{\bar{x}y} P'$, then $x, y \in$ {\myfont fn($P$)}.
\end{lemma}

\begin{lemma} \label{lm22}
If $P \xrightarrow{x(y)} P'$, then $x \in$ {\myfont fn($P$)}.
\end{lemma}

\begin{lemma} \label{lm23}
If $P \xrightarrow{\bar{x}(z)} P'$, then $x \in$ {\myfont fn($P$)} and $z \in$ {\myfont bn($P$)}.
\end{lemma}

\begin{lemma} \label{lm24}
If $P \xrightarrow{\alpha} P'$, $x \notin$ {\myfont n($\alpha$)} and $x \notin$ {\myfont fn($P$)}, then $x \notin$  {\myfont fn($P'$)}.
\end{lemma}

\begin{lemma} \label{lm25}
If $P \equiv P'$, then $x \in$ {\myfont fn$(P)$} $\Leftrightarrow$ $x \in$ {\myfont fn$(P')$.}
\end{lemma}

The first four lemmas follow by induction over the structure of the given transition. The last 
 by induction on the congruence judgment.

 The next key ingredient is establishing that structural congruence is a
strong late bisimulation.
 \begin{lemma} \label{lm26}
\marginnote{\href{https://github.com/GabrieleCecilia/concurrent-benchmark-solution/blob/main/code/6_stepcong_lemma.bel}{\ExternalLink}}
Let $P \equiv Q$.
\begin{enumerate}
	\item If $P \xrightarrow{\alpha} P'$, then there exists a process $Q'$ such that $Q \xrightarrow{\alpha} Q'$ and $P' \equiv Q'$.
	\item If $Q \xrightarrow{\alpha} Q'$, then there exists a process $P'$ such that $P \xrightarrow{\alpha} P'$ and $P' \equiv Q'$.
\end{enumerate}
\end{lemma}
The two statements need to be proven at the
same time by mutual induction over the derivation of the
congruence judgment and case analysis on the given transition.

Finally, a rewriting lemma for reduction, again proven by induction on
the structure of the given reduction judgment:
\begin{lemma} \label{lm27}
\marginnote{\href{https://github.com/GabrieleCecilia/concurrent-benchmark-solution/blob/main/code/7_reduction_rewriting.bel}{\ExternalLink}}
If $P \rightarrow Q$ then there exist three
  names $x,y$ and $z$, a finite (possibly empty) set of names
  $w_1,\ldots, w_n$ and three processes $R_1, R_2$ and $S$ such that 
  $P \equiv (\nu w_1)\ldots (\nu w_n) (\,(\bar{x}y.R_1 \mid x(z).R_2)
  \mid S\,)$ and 
  $Q \equiv (\nu w_1)\ldots (\nu w_n) (\,(R_1 \mid R_2 \{ y/z \}) \mid
  S\,)$.
\end{lemma}

\addtocounter{theorem}{-1}

\begin{theorem} \label{thm2}
\marginnote{\href{https://github.com/GabrieleCecilia/concurrent-benchmark-solution/blob/main/code/8_theorem2.bel}{\ExternalLink}}
$P \rightarrow Q$ implies the existence of a $Q'$ such that $P \xrightarrow{\tau} Q'$ and $Q \equiv Q'$.
\end{theorem}

The proof follows immediately from the application of
Lemmas~\ref{lm26} and~\ref{lm27}.


\section{Beluga Formalization}\label{sec:form}

This section provides an overview of the formalization of the
definitions and proofs introduced in the previous section with the
proof assistant Beluga. The complete formalization is accessible
at~\cite{GBThesis}.

\subsection{Syntax}
Fig.~\ref{fig:proc} displays the syntax of names and processes.
Since names are just an infinite set, we encode them with an LF type
{\ttfamily \small \color{belugagreen}names} without any constructor,
which will be extended with new inhabitants dynamically in the operational semantics.
This is made possible by the declaration
\begin{lstlisting}[linerange={Contexts-End}]
%%%%%%%%%%%%%%%%%%%%%%%%%%%%%%%%%%%%%%%%%%%%%%%%%%%%%%%%%

%%% Definitions %%%

%% Names %%
LF names: type =
;
%% End %%

%% Processes %%
LF proc: type =
  | p_zero: proc
  | p_in: names → (names → proc) → proc
  | p_out: names → names → proc → proc
  | p_par: proc → proc → proc ---infix p_par 11 left.
  | p_res: (names → proc) → proc
;
%% End %%

%% Contexts %%
schema ctx = names;
%% End %%

%%%%%%%%%%%%%%%%%%%%%%%%%%%%%%%%%%%%%%%%%%%%%%%%%%%%%%%%%

%%% Reduction Semantics %%%

%% Cong %%
% Structural Congruence
LF cong: proc → proc → type =
% Abelian Monoid Laws for Parallel Composition
  | par_assoc: cong (P p_par (Q p_par R)) ((P p_par Q) p_par R)
  | par_unit: cong (P p_par p_zero) P
  | par_comm: cong (P p_par Q) (Q p_par P)
% Scope Extension Laws
  | sc_ext_zero: cong (p_res (\x.p_zero)) p_zero
  | sc_ext_par: cong ((p_res P) p_par Q) (p_res (\x.((P x) p_par Q)))
  | sc_ext_res: cong (p_res \x.(p_res \y.(P x y))) (p_res \y.(p_res \x.(P x y)))
% Compatibility Laws
  | c_in: ({y:names} cong (P y) (Q y)) → cong (p_in X P) (p_in X Q)
  | c_out: cong P Q → cong (p_out X Y P) (p_out X Y Q)
  | c_par: cong P P' → cong (P p_par Q) (P' p_par Q)
  | c_res: ({x:names} cong (P x) (Q x)) → cong (p_res P) (p_res Q)
% Equivalence Relation Laws
  | c_ref: cong P P
  | c_sym: cong P Q → cong Q P
  | c_trans: cong P Q → cong Q R → cong P R
%% End %%
;
--infix cong 11 left.

%% Red %%

% Reduction
LF red: proc → proc → type =
  | r_com: red ((p_out X Y P) p_par (p_in X Q)) (P p_par (Q Y))
  | r_par: red P Q → red (P p_par R) (Q p_par R)
  | r_res: ({x:names} red (P x) (Q x)) → red (p_res P) (p_res Q)
  | r_str: P cong P' → red P' Q' → Q' cong Q → red P Q
%% End %%
;
--infix red 11 left.


%%%%%%%%%%%%%%%%%%%%%%%%%%%%%%%%%%%%%%%%%%%%%%%%%%%%%%%%%

%%% LTS Semantics %%%
% We follow "Pi-Calculus in (Co)Inductive Type Theory" [Honsell et al. 2001] for the encoding of late LTS semantics.
% We define two different types for free/bound actions, and two different relations for transitions via free/bound actions.
% The result of a free transition is a process, while the result of a bound transition is a function from names to processes:
% instead of stating the bound name involved in the transition explicitly, that name is the argument of the aforementioned function.

%% Act %%
% Free Actions                             % Bound Actions
LF f_act: type =                           LF b_act: type =
  | f_tau: f_act                             | b_in: names → b_act
  | f_out: names → names → f_act           | b_out: names → b_act
%% End %%
  ;                                          ;

% Bound Actions
LF b_act: type =
  | b_in: names → b_act
  | b_out: names → b_act
;

%% Trans %%

% Transition Relation
LF fstep: proc → f_act → proc → type =
  | fs_out: fstep (p_out X Y P) (f_out X Y) P
  | fs_par1: fstep P A P' → fstep (P p_par Q) A (P' p_par Q)
  | fs_par2: fstep Q A Q' → fstep (P p_par Q) A (P p_par Q')
  | fs_com1: fstep P (f_out X Y) P' → bstep Q (b_in X) Q'
    → fstep (P p_par Q) f_tau (P' p_par (Q' Y))
  | fs_com2: bstep P (b_in X) P' → fstep Q (f_out X Y) Q'
    → fstep (P p_par Q) f_tau ((P' Y) p_par Q')
  | fs_res: ({z:names} fstep (P z) A (P' z)) → fstep (p_res P) A (p_res P')
  | fs_close1: bstep P (b_out X) P' → bstep Q (b_in X) Q'
    → fstep (P p_par Q) f_tau (p_res \z.((P' z) p_par (Q' z)))
  | fs_close2: bstep P (b_in X) P' → bstep Q (b_out X) Q'
    → fstep (P p_par Q) f_tau (p_res \z.((P' z) p_par (Q' z)))
	    
and bstep: proc → b_act → (names → proc) → type =
  | bs_in: bstep (p_in X P) (b_in X) P
  | bs_par1: bstep P A P' → bstep (P p_par Q) A \x.((P' x) p_par Q)
  | bs_par2: bstep Q A Q' → bstep (P p_par Q) A \x.(P p_par (Q' x))
  | bs_res: ({z:names} bstep (P z) A (P' z))
    → bstep (p_res P) A \x.(p_res \z.(P' z x))
  | bs_open: ({z:names} fstep (P z) (f_out X z) (P' z)) → bstep (p_res P)(b_out X) P'
%% End %%
;
\end{lstlisting}
indicating that all relevant judgments involving open terms (processes) are formulated
 in contexts categorized by the above schema.

Processes are encoded using (weak) HOAS, as well trodden in the
literature: input and restrictions abstract over names, in particular
the input process $x(y).P$ is encoded by the term {\small
  \verb|p_in X \y.(P y)|}, where the bound name $y$ in $P$ is
represented by the implicit argument of the LF function {\small
  \verb|\y.(P y)|}.  As usual, $\alpha$-renaming and capture-avoiding
substitutions are automatically implemented by the meta-language. From
now on, semicolons and infix constructor declarations will be omitted
from code snippets for brevity.
 

\begin{figure}[th]
\begin{lstlisting}
\end{lstlisting}
\begin{lstlisting}[style=belugatwo,linerange={Names-End,Processes-End}]
%%%%%%%%%%%%%%%%%%%%%%%%%%%%%%%%%%%%%%%%%%%%%%%%%%%%%%%%%

%%% Definitions %%%

%% Names %%
LF names: type =
;
%% End %%

%% Processes %%
LF proc: type =
  | p_zero: proc
  | p_in: names → (names → proc) → proc
  | p_out: names → names → proc → proc
  | p_par: proc → proc → proc ---infix p_par 11 left.
  | p_res: (names → proc) → proc
;
%% End %%

%% Contexts %%
schema ctx = names;
%% End %%

%%%%%%%%%%%%%%%%%%%%%%%%%%%%%%%%%%%%%%%%%%%%%%%%%%%%%%%%%

%%% Reduction Semantics %%%

%% Cong %%
% Structural Congruence
LF cong: proc → proc → type =
% Abelian Monoid Laws for Parallel Composition
  | par_assoc: cong (P p_par (Q p_par R)) ((P p_par Q) p_par R)
  | par_unit: cong (P p_par p_zero) P
  | par_comm: cong (P p_par Q) (Q p_par P)
% Scope Extension Laws
  | sc_ext_zero: cong (p_res (\x.p_zero)) p_zero
  | sc_ext_par: cong ((p_res P) p_par Q) (p_res (\x.((P x) p_par Q)))
  | sc_ext_res: cong (p_res \x.(p_res \y.(P x y))) (p_res \y.(p_res \x.(P x y)))
% Compatibility Laws
  | c_in: ({y:names} cong (P y) (Q y)) → cong (p_in X P) (p_in X Q)
  | c_out: cong P Q → cong (p_out X Y P) (p_out X Y Q)
  | c_par: cong P P' → cong (P p_par Q) (P' p_par Q)
  | c_res: ({x:names} cong (P x) (Q x)) → cong (p_res P) (p_res Q)
% Equivalence Relation Laws
  | c_ref: cong P P
  | c_sym: cong P Q → cong Q P
  | c_trans: cong P Q → cong Q R → cong P R
%% End %%
;
--infix cong 11 left.

%% Red %%

% Reduction
LF red: proc → proc → type =
  | r_com: red ((p_out X Y P) p_par (p_in X Q)) (P p_par (Q Y))
  | r_par: red P Q → red (P p_par R) (Q p_par R)
  | r_res: ({x:names} red (P x) (Q x)) → red (p_res P) (p_res Q)
  | r_str: P cong P' → red P' Q' → Q' cong Q → red P Q
%% End %%
;
--infix red 11 left.


%%%%%%%%%%%%%%%%%%%%%%%%%%%%%%%%%%%%%%%%%%%%%%%%%%%%%%%%%

%%% LTS Semantics %%%
% We follow "Pi-Calculus in (Co)Inductive Type Theory" [Honsell et al. 2001] for the encoding of late LTS semantics.
% We define two different types for free/bound actions, and two different relations for transitions via free/bound actions.
% The result of a free transition is a process, while the result of a bound transition is a function from names to processes:
% instead of stating the bound name involved in the transition explicitly, that name is the argument of the aforementioned function.

%% Act %%
% Free Actions                             % Bound Actions
LF f_act: type =                           LF b_act: type =
  | f_tau: f_act                             | b_in: names → b_act
  | f_out: names → names → f_act           | b_out: names → b_act
%% End %%
  ;                                          ;

% Bound Actions
LF b_act: type =
  | b_in: names → b_act
  | b_out: names → b_act
;

%% Trans %%

% Transition Relation
LF fstep: proc → f_act → proc → type =
  | fs_out: fstep (p_out X Y P) (f_out X Y) P
  | fs_par1: fstep P A P' → fstep (P p_par Q) A (P' p_par Q)
  | fs_par2: fstep Q A Q' → fstep (P p_par Q) A (P p_par Q')
  | fs_com1: fstep P (f_out X Y) P' → bstep Q (b_in X) Q'
    → fstep (P p_par Q) f_tau (P' p_par (Q' Y))
  | fs_com2: bstep P (b_in X) P' → fstep Q (f_out X Y) Q'
    → fstep (P p_par Q) f_tau ((P' Y) p_par Q')
  | fs_res: ({z:names} fstep (P z) A (P' z)) → fstep (p_res P) A (p_res P')
  | fs_close1: bstep P (b_out X) P' → bstep Q (b_in X) Q'
    → fstep (P p_par Q) f_tau (p_res \z.((P' z) p_par (Q' z)))
  | fs_close2: bstep P (b_in X) P' → bstep Q (b_out X) Q'
    → fstep (P p_par Q) f_tau (p_res \z.((P' z) p_par (Q' z)))
	    
and bstep: proc → b_act → (names → proc) → type =
  | bs_in: bstep (p_in X P) (b_in X) P
  | bs_par1: bstep P A P' → bstep (P p_par Q) A \x.((P' x) p_par Q)
  | bs_par2: bstep Q A Q' → bstep (P p_par Q) A \x.(P p_par (Q' x))
  | bs_res: ({z:names} bstep (P z) A (P' z))
    → bstep (p_res P) A \x.(p_res \z.(P' z x))
  | bs_open: ({z:names} fstep (P z) (f_out X z) (P' z)) → bstep (p_res P)(b_out X) P'
%% End %%
;
\end{lstlisting}
\vspace{-0.5\baselineskip}
\caption{Encoding of names and processes.}
\label{fig:proc}
\end{figure}

Recall that LF types are \emph{not} inductive.  While the LF type
{\ttfamily \small \color{belugagreen}names} has no constructor, the
\emph{contextual} type {\small \color{belugagreen}
  \verb|[g |$\vdash$\verb| proc]|}, for {\small \verb|g|} $\in$ {\ttfamily \small
  \color{belugagreen} ctx}, denotes the set of open processes built out
of LF variables ranging over names.

\subsection{Reduction Semantics}

Congruence and reduction are encoded by the type families
{\ttfamily \small \color{belugagreen}cong} and {\ttfamily \small
  \color{belugagreen}red} respectively, as presented in
Fig.~\ref{fig:belcongred}. As usual, we use universal quantification
{\small \verb|{x:names}|} to descend into binders, e.g.\ in
the compatibility rule for restriction. Scope extension is realized in rule
{\small\verb|sc_ext_par|} by
simply \emph{not} having $Q$ depend on the restricted channel, hence
meeting the side condition $x \notin$ {\myfont fn$(Q)$} in the {\small
  \textsc{Sc-Ext-Par}}
automatically. 
In the same vein, substitution in rule {\small \textsc{R-Com}} is encoded by meta-level $\beta$-reduction.
\begin{figure}[h]
\begin{lstlisting}[style=belugatwo,linerange={Cong-End,Red-End}]
%%%%%%%%%%%%%%%%%%%%%%%%%%%%%%%%%%%%%%%%%%%%%%%%%%%%%%%%%

%%% Definitions %%%

%% Names %%
LF names: type =
;
%% End %%

%% Processes %%
LF proc: type =
  | p_zero: proc
  | p_in: names → (names → proc) → proc
  | p_out: names → names → proc → proc
  | p_par: proc → proc → proc ---infix p_par 11 left.
  | p_res: (names → proc) → proc
;
%% End %%

%% Contexts %%
schema ctx = names;
%% End %%

%%%%%%%%%%%%%%%%%%%%%%%%%%%%%%%%%%%%%%%%%%%%%%%%%%%%%%%%%

%%% Reduction Semantics %%%

%% Cong %%
% Structural Congruence
LF cong: proc → proc → type =
% Abelian Monoid Laws for Parallel Composition
  | par_assoc: cong (P p_par (Q p_par R)) ((P p_par Q) p_par R)
  | par_unit: cong (P p_par p_zero) P
  | par_comm: cong (P p_par Q) (Q p_par P)
% Scope Extension Laws
  | sc_ext_zero: cong (p_res (\x.p_zero)) p_zero
  | sc_ext_par: cong ((p_res P) p_par Q) (p_res (\x.((P x) p_par Q)))
  | sc_ext_res: cong (p_res \x.(p_res \y.(P x y))) (p_res \y.(p_res \x.(P x y)))
% Compatibility Laws
  | c_in: ({y:names} cong (P y) (Q y)) → cong (p_in X P) (p_in X Q)
  | c_out: cong P Q → cong (p_out X Y P) (p_out X Y Q)
  | c_par: cong P P' → cong (P p_par Q) (P' p_par Q)
  | c_res: ({x:names} cong (P x) (Q x)) → cong (p_res P) (p_res Q)
% Equivalence Relation Laws
  | c_ref: cong P P
  | c_sym: cong P Q → cong Q P
  | c_trans: cong P Q → cong Q R → cong P R
%% End %%
;
--infix cong 11 left.

%% Red %%

% Reduction
LF red: proc → proc → type =
  | r_com: red ((p_out X Y P) p_par (p_in X Q)) (P p_par (Q Y))
  | r_par: red P Q → red (P p_par R) (Q p_par R)
  | r_res: ({x:names} red (P x) (Q x)) → red (p_res P) (p_res Q)
  | r_str: P cong P' → red P' Q' → Q' cong Q → red P Q
%% End %%
;
--infix red 11 left.


%%%%%%%%%%%%%%%%%%%%%%%%%%%%%%%%%%%%%%%%%%%%%%%%%%%%%%%%%

%%% LTS Semantics %%%
% We follow "Pi-Calculus in (Co)Inductive Type Theory" [Honsell et al. 2001] for the encoding of late LTS semantics.
% We define two different types for free/bound actions, and two different relations for transitions via free/bound actions.
% The result of a free transition is a process, while the result of a bound transition is a function from names to processes:
% instead of stating the bound name involved in the transition explicitly, that name is the argument of the aforementioned function.

%% Act %%
% Free Actions                             % Bound Actions
LF f_act: type =                           LF b_act: type =
  | f_tau: f_act                             | b_in: names → b_act
  | f_out: names → names → f_act           | b_out: names → b_act
%% End %%
  ;                                          ;

% Bound Actions
LF b_act: type =
  | b_in: names → b_act
  | b_out: names → b_act
;

%% Trans %%

% Transition Relation
LF fstep: proc → f_act → proc → type =
  | fs_out: fstep (p_out X Y P) (f_out X Y) P
  | fs_par1: fstep P A P' → fstep (P p_par Q) A (P' p_par Q)
  | fs_par2: fstep Q A Q' → fstep (P p_par Q) A (P p_par Q')
  | fs_com1: fstep P (f_out X Y) P' → bstep Q (b_in X) Q'
    → fstep (P p_par Q) f_tau (P' p_par (Q' Y))
  | fs_com2: bstep P (b_in X) P' → fstep Q (f_out X Y) Q'
    → fstep (P p_par Q) f_tau ((P' Y) p_par Q')
  | fs_res: ({z:names} fstep (P z) A (P' z)) → fstep (p_res P) A (p_res P')
  | fs_close1: bstep P (b_out X) P' → bstep Q (b_in X) Q'
    → fstep (P p_par Q) f_tau (p_res \z.((P' z) p_par (Q' z)))
  | fs_close2: bstep P (b_in X) P' → bstep Q (b_out X) Q'
    → fstep (P p_par Q) f_tau (p_res \z.((P' z) p_par (Q' z)))
	    
and bstep: proc → b_act → (names → proc) → type =
  | bs_in: bstep (p_in X P) (b_in X) P
  | bs_par1: bstep P A P' → bstep (P p_par Q) A \x.((P' x) p_par Q)
  | bs_par2: bstep Q A Q' → bstep (P p_par Q) A \x.(P p_par (Q' x))
  | bs_res: ({z:names} bstep (P z) A (P' z))
    → bstep (p_res P) A \x.(p_res \z.(P' z x))
  | bs_open: ({z:names} fstep (P z) (f_out X z) (P' z)) → bstep (p_res P)(b_out X) P'
%% End %%
;
\end{lstlisting}
\vspace{-0.5\baselineskip}
\caption{Encoding of congruence and reduction.}
\label{fig:belcongred}
\end{figure}

\subsection{Labelled Transition System Semantics}
We follow Honsell et al.~\cite{DBLP:journals/tcs/HonsellMS01} for the
encoding of the late LTS semantics. We declare two
different relations for transitions via free and bound actions. The
result of a free transition is a process, while the result of a bound
transition is a process abstraction: instead of explicitly stating the
bound name involved in the transition, that name is the argument of
the aforementioned function. This is reflected by the encoding of
free and bound actions, which only mention the free names involved.
 Fig.~\ref{fig:acttrans} shows the types {\small
  \color{belugagreen}\verb|f_act|} and {\small
  \color{belugagreen}\verb|b_act|} encoding free and bound
actions and the  two mutually defined
type families {\ttfamily \small \color{belugagreen}fstep} and
{\ttfamily \small \color{belugagreen}bstep} which encode free and bound transitions
respectively\footnote{Interestingly, a similar approach is pursued by Cheney in his nominal encoding in $\alpha$Prolog, see\\ \url{https://homepages.inf.ed.ac.uk/jcheney/programs/aprolog/examples/picalc.apl}.}. Note that none of the side conditions
of the transition rules need to be explicitly stated, nor do we need the axioms and additional freshness predicates as in~\cite{DBLP:journals/tcs/HonsellMS01}.

HOAS encodings customarily come with an (informal) \emph{adequacy}
proof, ensuring that there is a compositional bijection between the
mathematical model and its encoding (in canonical form). While this is
fairly obvious for processes, congruence and reduction, it is less so
w.r.t.\ the LTS semantics. Luckily, this has been carefully proven
both in~\cite{DBLP:journals/tcs/HonsellMS01} and
in~\cite{DBLP:journals/tocl/TiuM10} for a related version.  We refer
to those papers for further details and to the repository for a Beluga proof of the ``internal'' adequacy of those two encodings.

\begin{figure}[h]
\begin{lstlisting}[style=belugatwo,linerange={Act-End,Trans-End}]
%%%%%%%%%%%%%%%%%%%%%%%%%%%%%%%%%%%%%%%%%%%%%%%%%%%%%%%%%

%%% Definitions %%%

%% Names %%
LF names: type =
;
%% End %%

%% Processes %%
LF proc: type =
  | p_zero: proc
  | p_in: names → (names → proc) → proc
  | p_out: names → names → proc → proc
  | p_par: proc → proc → proc ---infix p_par 11 left.
  | p_res: (names → proc) → proc
;
%% End %%

%% Contexts %%
schema ctx = names;
%% End %%

%%%%%%%%%%%%%%%%%%%%%%%%%%%%%%%%%%%%%%%%%%%%%%%%%%%%%%%%%

%%% Reduction Semantics %%%

%% Cong %%
% Structural Congruence
LF cong: proc → proc → type =
% Abelian Monoid Laws for Parallel Composition
  | par_assoc: cong (P p_par (Q p_par R)) ((P p_par Q) p_par R)
  | par_unit: cong (P p_par p_zero) P
  | par_comm: cong (P p_par Q) (Q p_par P)
% Scope Extension Laws
  | sc_ext_zero: cong (p_res (\x.p_zero)) p_zero
  | sc_ext_par: cong ((p_res P) p_par Q) (p_res (\x.((P x) p_par Q)))
  | sc_ext_res: cong (p_res \x.(p_res \y.(P x y))) (p_res \y.(p_res \x.(P x y)))
% Compatibility Laws
  | c_in: ({y:names} cong (P y) (Q y)) → cong (p_in X P) (p_in X Q)
  | c_out: cong P Q → cong (p_out X Y P) (p_out X Y Q)
  | c_par: cong P P' → cong (P p_par Q) (P' p_par Q)
  | c_res: ({x:names} cong (P x) (Q x)) → cong (p_res P) (p_res Q)
% Equivalence Relation Laws
  | c_ref: cong P P
  | c_sym: cong P Q → cong Q P
  | c_trans: cong P Q → cong Q R → cong P R
%% End %%
;
--infix cong 11 left.

%% Red %%

% Reduction
LF red: proc → proc → type =
  | r_com: red ((p_out X Y P) p_par (p_in X Q)) (P p_par (Q Y))
  | r_par: red P Q → red (P p_par R) (Q p_par R)
  | r_res: ({x:names} red (P x) (Q x)) → red (p_res P) (p_res Q)
  | r_str: P cong P' → red P' Q' → Q' cong Q → red P Q
%% End %%
;
--infix red 11 left.


%%%%%%%%%%%%%%%%%%%%%%%%%%%%%%%%%%%%%%%%%%%%%%%%%%%%%%%%%

%%% LTS Semantics %%%
% We follow "Pi-Calculus in (Co)Inductive Type Theory" [Honsell et al. 2001] for the encoding of late LTS semantics.
% We define two different types for free/bound actions, and two different relations for transitions via free/bound actions.
% The result of a free transition is a process, while the result of a bound transition is a function from names to processes:
% instead of stating the bound name involved in the transition explicitly, that name is the argument of the aforementioned function.

%% Act %%
% Free Actions                             % Bound Actions
LF f_act: type =                           LF b_act: type =
  | f_tau: f_act                             | b_in: names → b_act
  | f_out: names → names → f_act           | b_out: names → b_act
%% End %%
  ;                                          ;

% Bound Actions
LF b_act: type =
  | b_in: names → b_act
  | b_out: names → b_act
;

%% Trans %%

% Transition Relation
LF fstep: proc → f_act → proc → type =
  | fs_out: fstep (p_out X Y P) (f_out X Y) P
  | fs_par1: fstep P A P' → fstep (P p_par Q) A (P' p_par Q)
  | fs_par2: fstep Q A Q' → fstep (P p_par Q) A (P p_par Q')
  | fs_com1: fstep P (f_out X Y) P' → bstep Q (b_in X) Q'
    → fstep (P p_par Q) f_tau (P' p_par (Q' Y))
  | fs_com2: bstep P (b_in X) P' → fstep Q (f_out X Y) Q'
    → fstep (P p_par Q) f_tau ((P' Y) p_par Q')
  | fs_res: ({z:names} fstep (P z) A (P' z)) → fstep (p_res P) A (p_res P')
  | fs_close1: bstep P (b_out X) P' → bstep Q (b_in X) Q'
    → fstep (P p_par Q) f_tau (p_res \z.((P' z) p_par (Q' z)))
  | fs_close2: bstep P (b_in X) P' → bstep Q (b_out X) Q'
    → fstep (P p_par Q) f_tau (p_res \z.((P' z) p_par (Q' z)))
	    
and bstep: proc → b_act → (names → proc) → type =
  | bs_in: bstep (p_in X P) (b_in X) P
  | bs_par1: bstep P A P' → bstep (P p_par Q) A \x.((P' x) p_par Q)
  | bs_par2: bstep Q A Q' → bstep (P p_par Q) A \x.(P p_par (Q' x))
  | bs_res: ({z:names} bstep (P z) A (P' z))
    → bstep (p_res P) A \x.(p_res \z.(P' z x))
  | bs_open: ({z:names} fstep (P z) (f_out X z) (P' z)) → bstep (p_res P)(b_out X) P'
%% End %%
;
\end{lstlisting}
\vspace{-0.5\baselineskip}
\caption{Encoding of actions and transition.}
\label{fig:acttrans}
\end{figure}

\subsection{The Harmony Lemma}

One of the expected yet  very much appreciated payoffs of a HOAS
encoding is that most (in fact all but Lemma~\ref{lm24}) boilerplate lemmas about
names, occurrences and substitution vanish. We are referring to the
substitution Lemmas~\ref{le:s1}---\ref{le:s3}, as well as the
free/bound names Lemmas~\ref{lm21},~\ref{lm22},~\ref{lm23}
and~\ref{lm25}.

\subsubsection{Theorem~\ref{thm1}: $\tau$-Transition Implies Reduction}
We start with the encoding of Lemmas~\ref{lm11},~\ref{lm12} and~\ref{lm13}. There are two issues, one standard, the other slightly
more challenging. For one, the conclusion of these lemmas includes an
existential quantification, and Beluga lacks  such a construct; the
usual workaround consists of defining a new type family which encodes
the existential quantification. Secondly, and more seriously, the statements refer to
sequences of binders (here restrictions), sometimes referred to as
\emph{telescopes}.
We can give a combined solution to these items by encoding them
\emph{inductively}, where the base case describes when the existential
holds for the empty sequence and the inductive one adds one more binder. 
More specifically for Lemma~\ref{lm11}, we say that
the congruences ($\star$)
($Q \equiv (\nu w_1)\ldots (\nu w_n) (x(y).R \mid S)$ and
\mbox{$Q' \equiv (\nu w_1)\ldots (\nu w_n) (R \mid S)$}) hold for two
processes $Q$ and $Q'$ iff one of the following holds:
\begin{enumerate}
	\item[i.] $Q \equiv x(y).R \mid S$ and $Q' \equiv R \mid S$;
	\item[ii.] $Q \equiv (\nu w)P$, $Q' \equiv (\nu w)P'$ and the congruences ($\star$) hold for $P$ and $P'$.
\end{enumerate}
        
We list the definition of the type family {\small
  \color{belugagreen}\verb|ex_inp_rew|} that encodes the above judgment; the types {\small
  \color{belugagreen}\verb|ex_fout_rew|} and {\small
  \color{belugagreen}\verb|ex_bout_rew|} are defined analogously.
\begin{lstlisting}[linerange={ExInpRew-End}]
% Predicate about syntactic equivalence of processes involved in input transitions:
% ex_inp_rew Q X Q' holds iff there exist R, S, x_1,...,x_n such that Q cong (nu x_1)...(nu x_n)((p_in X R) p_par S) and
% (Q' y) cong (nu x_1)...(nu x_n)((R Y) p_par S).
% We encode this type via two constructors: one representing the base case without restrictions, and one representing the inductive case
% in which Q cong (nu x)P and Q' cong (nu x)P', where ex_inp_rew P X P' holds inductively.

%% ExInpRew %%
LF ex_inp_rew: proc → names → (names → proc) → type =
  | inp_base: Q cong ((p_in X R) p_par S) → ({y:names} (Q' y) cong ((R y) p_par S))
    → ex_inp_rew Q X Q'
  | inp_ind: Q cong (p_res P) → ({y:names} (Q' y) cong (p_res (P' y)))
    → ({w:names} ex_inp_rew (P w) X \y.(P' y w)) → ex_inp_rew Q X Q'
%% End %%
;


% We prove the lemma which states that processes involved in input transitions (bstep Q (b_in X) Q')
% can be rewritten (up to congruence) in a canonic way (ex_inp_rew Q X Q').
% Before, we need three additional lemmas which prove the result in particular cases.

%% BsInRewPar1 %%
rec bs_in_rew_par1: (g:ctx) {R:[g ⊢ proc]} [g ⊢ ex_inp_rew Q X \y.Q'[..,y]]
  → [g ⊢ ex_inp_rew (Q p_par R) X \y.(Q'[..,y] p_par R[..])] = ...
%% End %%
/ total d (bs_in_rew_par1 _ _ _ _ _ d) /
mlam R ⇒ fn d ⇒ case d of
   | [g ⊢ inp_base C1 \y.C2[..,y]] ⇒ [g ⊢ inp_base (c_trans (c_par C1)
     (c_sym par_assoc)) \y.(c_trans (c_par C2[..,y]) (c_sym par_assoc))]
   | [g ⊢ inp_ind C1 (\y.C2[..,y]) (\w.D1[..,w])] ⇒
     let [g,w:names ⊢ D2[..,w]] = bs_in_rew_par1 [g,w:names ⊢ R[..]]
     [g,w:names ⊢ D1[..,w]] in [g ⊢ inp_ind (c_trans (c_par C1)
     sc_ext_par) (\y.(c_trans (c_par C2[..,y]) sc_ext_par)) (\w.D2[..,w])]
;

rec bs_in_rew_par2:(g:ctx){R:[g ⊢ proc]}[g ⊢ ex_inp_rew Q X \y.Q'[..,y]] → [g ⊢ ex_inp_rew (R p_par Q) X \y.(R[..] p_par Q'[..,y])] = ?
;
%{mlam R ⇒ fn d ⇒ case d of
  | [g ⊢ inp_base C1 \y.C2[..,y]] ⇒ [g ⊢ inp_base (c_trans par_comm (c_trans (c_par C1) (c_sym par_assoc))) \y.(c_trans par_comm (c_trans (c_par C2[..,y]) (c_sym par_assoc)))]
  | [g ⊢ inp_ind C1 (\y.C2[..,y]) (\w.D1[..,w])] ⇒ let [g,w:names ⊢ D2[..,w]] = bs_in_rew_par2 [g,w:names ⊢ R[..]] [g,w:names ⊢ D1[..,w]]
    in [g ⊢ inp_ind (c_trans par_comm (c_trans (c_par C1) (c_trans sc_ext_par (c_res \w.par_comm)))) (\y.(c_trans par_comm (c_trans (c_par C2[..,y]) (c_trans sc_ext_par (c_res \w.par_comm))))) (\w.D2[..,w])]
;}%

rec bs_in_rew_res: (g:ctx) [g,z:names ⊢ ex_inp_rew Q[..,z] X[..] \y.Q'[..,z,y]] → [g ⊢ ex_inp_rew (p_res \z.Q[..,z]) X \y.(p_res \z.Q'[..,z,y])] = ?
;
%{fn d ⇒ case d of
  | [g,z:names ⊢ inp_base C1[..,z] \y.C2[..,z,y]] ⇒ [g ⊢ inp_ind (c_res \z.C1[..,z]) (\y.(c_res \z.C2[..,z,y])) (\z.(inp_base c_ref \y.c_ref))]
  | [g,z:names ⊢ inp_ind C1[..,z] (\y.C2[..,z,y]) (\w.D1[..,z,w])] ⇒ let [g,z:names ⊢ D2[..,z]] = bs_in_rew_res [g,z:names,w:names ⊢
    D1[..,z,w]] in [g ⊢ inp_ind (c_res \z.C1[..,z]) (\y.(c_res \z.C2[..,z,y])) (\z.D2[..,z])]
;}%

%% BsInRew %%
rec bs_in_rew: (g:ctx) [g ⊢ bstep Q (b_in X) \y.Q'[..,y]]
  → [g ⊢ ex_inp_rew Q X \y.Q'[..,y]] =
/ total b (bs_in_rew _ _ _ _ b) /
fn b ⇒ case b of
   | [g ⊢ bs_in] ⇒ [g ⊢ inp_base (c_sym par_unit) \y.(c_sym par_unit)]
   | [g ⊢ bs_par1 B1]:[g ⊢ bstep (P p_par R) (b_in X) (\y.(P' p_par (R[..])))] ⇒
     let [g ⊢ D1] = bs_in_rew [g ⊢ B1] in
     let [g ⊢ D2] = bs_in_rew_par1 [g ⊢ R] [g ⊢ D1] in [g ⊢ D2]
   | [g ⊢ bs_par2 B2]:[g ⊢ bstep (R p_par P) (b_in X) (\y.((R[..]) p_par P'))] ⇒
     let [g ⊢ D1] = bs_in_rew [g ⊢ B2] in 
     let [g ⊢ D2] = bs_in_rew_par2 [g ⊢ R] [g ⊢ D1] in [g ⊢ D2]
   | [g ⊢ bs_res \y.B[..,y]] ⇒
     let [g,y:names ⊢ D1[..,y]] = bs_in_rew [g,y:names ⊢ B[..,y]] in
     let [g ⊢ D2] = bs_in_rew_res [g,y:names ⊢ D1[..,y]] in [g ⊢ D2]
%% End %%
;
%{
/ total q (bs_in_rew _ q _ _ _) /
mlam Q ⇒ fn b ⇒ case [_ ⊢ Q] of
   | [g ⊢ p_zero] ⇒ impossible b
   | [g ⊢ p_in W \z.P[..,z]] ⇒ let [g ⊢ bs_in] = b in
     [g ⊢ inp_base (c_sym par_unit) \y.(c_sym par_unit)]
   | [g ⊢ p_out W Z P] ⇒ impossible b
   | [g ⊢ Q1 p_par Q2] ⇒
     (case b of
        | [g ⊢ bs_par1 B1] ⇒
          let [g ⊢ D1] = bs_in_rew [g ⊢ Q1] [g ⊢ B1] in
          let [g ⊢ D2] = bs_in_rew_par1 [g ⊢ Q2] [g ⊢ D1] in [g ⊢ D2]
        | [g ⊢ bs_par2 B2] ⇒
          let [g ⊢ D1] = bs_in_rew [g ⊢ Q2] [g ⊢ B2] in
          let [g ⊢ D2] = bs_in_rew_par2 [g ⊢ Q1] [g ⊢ D1] in [g ⊢ D2]
      )
   | [g ⊢ p_res \z.P[..,z]] ⇒ let [g ⊢ bs_res \z.B[..,z]] = b in
     let [g,z:names ⊢ D1[..,z]] = bs_in_rew [g,z:names ⊢ P[..,z]]
     [g,z:names ⊢ B[..,z]] in
     let [g ⊢ D2] = bs_in_rew_res [g,z:names ⊢ D1[..,z]] in [g ⊢ D2]
}%
\end{lstlisting}

We prove each lemma by defining a total recursive function which
receives a contextual derivation of an input/free output/bound output
transition respectively and returns a contextual object of the
corresponding existential type. We provide the proof term of the {\small
  \color{belugablue} \verb|bs_in_rew|} function, which proves Lemma~\ref{lm11}, in Fig.~\ref{fig:sampleproof}; the functions encoding 
Lemmas~\ref{lm12} and~\ref{lm13} are omitted.
\begin{figure}
\begin{lstlisting}[style=belugatwo,linerange={BsInRew-End}]
% Predicate about syntactic equivalence of processes involved in input transitions:
% ex_inp_rew Q X Q' holds iff there exist R, S, x_1,...,x_n such that Q cong (nu x_1)...(nu x_n)((p_in X R) p_par S) and
% (Q' y) cong (nu x_1)...(nu x_n)((R Y) p_par S).
% We encode this type via two constructors: one representing the base case without restrictions, and one representing the inductive case
% in which Q cong (nu x)P and Q' cong (nu x)P', where ex_inp_rew P X P' holds inductively.

%% ExInpRew %%
LF ex_inp_rew: proc → names → (names → proc) → type =
  | inp_base: Q cong ((p_in X R) p_par S) → ({y:names} (Q' y) cong ((R y) p_par S))
    → ex_inp_rew Q X Q'
  | inp_ind: Q cong (p_res P) → ({y:names} (Q' y) cong (p_res (P' y)))
    → ({w:names} ex_inp_rew (P w) X \y.(P' y w)) → ex_inp_rew Q X Q'
%% End %%
;


% We prove the lemma which states that processes involved in input transitions (bstep Q (b_in X) Q')
% can be rewritten (up to congruence) in a canonic way (ex_inp_rew Q X Q').
% Before, we need three additional lemmas which prove the result in particular cases.

%% BsInRewPar1 %%
rec bs_in_rew_par1: (g:ctx) {R:[g ⊢ proc]} [g ⊢ ex_inp_rew Q X \y.Q'[..,y]]
  → [g ⊢ ex_inp_rew (Q p_par R) X \y.(Q'[..,y] p_par R[..])] = ...
%% End %%
/ total d (bs_in_rew_par1 _ _ _ _ _ d) /
mlam R ⇒ fn d ⇒ case d of
   | [g ⊢ inp_base C1 \y.C2[..,y]] ⇒ [g ⊢ inp_base (c_trans (c_par C1)
     (c_sym par_assoc)) \y.(c_trans (c_par C2[..,y]) (c_sym par_assoc))]
   | [g ⊢ inp_ind C1 (\y.C2[..,y]) (\w.D1[..,w])] ⇒
     let [g,w:names ⊢ D2[..,w]] = bs_in_rew_par1 [g,w:names ⊢ R[..]]
     [g,w:names ⊢ D1[..,w]] in [g ⊢ inp_ind (c_trans (c_par C1)
     sc_ext_par) (\y.(c_trans (c_par C2[..,y]) sc_ext_par)) (\w.D2[..,w])]
;

rec bs_in_rew_par2:(g:ctx){R:[g ⊢ proc]}[g ⊢ ex_inp_rew Q X \y.Q'[..,y]] → [g ⊢ ex_inp_rew (R p_par Q) X \y.(R[..] p_par Q'[..,y])] = ?
;
%{mlam R ⇒ fn d ⇒ case d of
  | [g ⊢ inp_base C1 \y.C2[..,y]] ⇒ [g ⊢ inp_base (c_trans par_comm (c_trans (c_par C1) (c_sym par_assoc))) \y.(c_trans par_comm (c_trans (c_par C2[..,y]) (c_sym par_assoc)))]
  | [g ⊢ inp_ind C1 (\y.C2[..,y]) (\w.D1[..,w])] ⇒ let [g,w:names ⊢ D2[..,w]] = bs_in_rew_par2 [g,w:names ⊢ R[..]] [g,w:names ⊢ D1[..,w]]
    in [g ⊢ inp_ind (c_trans par_comm (c_trans (c_par C1) (c_trans sc_ext_par (c_res \w.par_comm)))) (\y.(c_trans par_comm (c_trans (c_par C2[..,y]) (c_trans sc_ext_par (c_res \w.par_comm))))) (\w.D2[..,w])]
;}%

rec bs_in_rew_res: (g:ctx) [g,z:names ⊢ ex_inp_rew Q[..,z] X[..] \y.Q'[..,z,y]] → [g ⊢ ex_inp_rew (p_res \z.Q[..,z]) X \y.(p_res \z.Q'[..,z,y])] = ?
;
%{fn d ⇒ case d of
  | [g,z:names ⊢ inp_base C1[..,z] \y.C2[..,z,y]] ⇒ [g ⊢ inp_ind (c_res \z.C1[..,z]) (\y.(c_res \z.C2[..,z,y])) (\z.(inp_base c_ref \y.c_ref))]
  | [g,z:names ⊢ inp_ind C1[..,z] (\y.C2[..,z,y]) (\w.D1[..,z,w])] ⇒ let [g,z:names ⊢ D2[..,z]] = bs_in_rew_res [g,z:names,w:names ⊢
    D1[..,z,w]] in [g ⊢ inp_ind (c_res \z.C1[..,z]) (\y.(c_res \z.C2[..,z,y])) (\z.D2[..,z])]
;}%

%% BsInRew %%
rec bs_in_rew: (g:ctx) [g ⊢ bstep Q (b_in X) \y.Q'[..,y]]
  → [g ⊢ ex_inp_rew Q X \y.Q'[..,y]] =
/ total b (bs_in_rew _ _ _ _ b) /
fn b ⇒ case b of
   | [g ⊢ bs_in] ⇒ [g ⊢ inp_base (c_sym par_unit) \y.(c_sym par_unit)]
   | [g ⊢ bs_par1 B1]:[g ⊢ bstep (P p_par R) (b_in X) (\y.(P' p_par (R[..])))] ⇒
     let [g ⊢ D1] = bs_in_rew [g ⊢ B1] in
     let [g ⊢ D2] = bs_in_rew_par1 [g ⊢ R] [g ⊢ D1] in [g ⊢ D2]
   | [g ⊢ bs_par2 B2]:[g ⊢ bstep (R p_par P) (b_in X) (\y.((R[..]) p_par P'))] ⇒
     let [g ⊢ D1] = bs_in_rew [g ⊢ B2] in 
     let [g ⊢ D2] = bs_in_rew_par2 [g ⊢ R] [g ⊢ D1] in [g ⊢ D2]
   | [g ⊢ bs_res \y.B[..,y]] ⇒
     let [g,y:names ⊢ D1[..,y]] = bs_in_rew [g,y:names ⊢ B[..,y]] in
     let [g ⊢ D2] = bs_in_rew_res [g,y:names ⊢ D1[..,y]] in [g ⊢ D2]
%% End %%
;
%{
/ total q (bs_in_rew _ q _ _ _) /
mlam Q ⇒ fn b ⇒ case [_ ⊢ Q] of
   | [g ⊢ p_zero] ⇒ impossible b
   | [g ⊢ p_in W \z.P[..,z]] ⇒ let [g ⊢ bs_in] = b in
     [g ⊢ inp_base (c_sym par_unit) \y.(c_sym par_unit)]
   | [g ⊢ p_out W Z P] ⇒ impossible b
   | [g ⊢ Q1 p_par Q2] ⇒
     (case b of
        | [g ⊢ bs_par1 B1] ⇒
          let [g ⊢ D1] = bs_in_rew [g ⊢ Q1] [g ⊢ B1] in
          let [g ⊢ D2] = bs_in_rew_par1 [g ⊢ Q2] [g ⊢ D1] in [g ⊢ D2]
        | [g ⊢ bs_par2 B2] ⇒
          let [g ⊢ D1] = bs_in_rew [g ⊢ Q2] [g ⊢ B2] in
          let [g ⊢ D2] = bs_in_rew_par2 [g ⊢ Q1] [g ⊢ D1] in [g ⊢ D2]
      )
   | [g ⊢ p_res \z.P[..,z]] ⇒ let [g ⊢ bs_res \z.B[..,z]] = b in
     let [g,z:names ⊢ D1[..,z]] = bs_in_rew [g,z:names ⊢ P[..,z]]
     [g,z:names ⊢ B[..,z]] in
     let [g ⊢ D2] = bs_in_rew_res [g,z:names ⊢ D1[..,z]] in [g ⊢ D2]
}%
\end{lstlisting}
\vspace{-0.5\baselineskip}
\caption{Proof of Lemma~\ref{lm11}.}
\label{fig:sampleproof}
\end{figure}

The proof proceeds by induction on the structure of the given assumption.  If
 it has been obtained through the {\small \textsc{S-In}} rule, we
are in the base case of the existential with no binders: hence, we
conclude immediately modulo symmetry of congruence. In case the input
transition has been obtained through the {\small \textsc{S-Par-L}}
rule, it can be rewritten as
$P \mid R \, \xrightarrow{x(y)} \, P' \mid R$, with the transition
{\small \verb|B1|}: $P \xrightarrow{x(y)} P'$ as hypothesis.  In such
situation we can apply a recursive call of the {\small
  \color{belugablue} \verb|bs_in_rew|} function to {\small \verb|B1|},
obtaining an object {\small \verb|D1|} of type
{\small\color{belugagreen}\verb|ex_inp_rew|} encoding the rewriting
for $P$ and $P'$; to conclude, we appeal to the lemma {\small
  \color{belugablue} \verb|bs_in_rew_par1|} in order to unfold {\small
  \verb|D1|} and build the required existential object encoding the
rewriting for $P \mid R$ and $P' \mid R$.  Here is the signature
of the above lemma:
\begin{lstlisting}[linerange={BsInRewPar1-End}]
% Predicate about syntactic equivalence of processes involved in input transitions:
% ex_inp_rew Q X Q' holds iff there exist R, S, x_1,...,x_n such that Q cong (nu x_1)...(nu x_n)((p_in X R) p_par S) and
% (Q' y) cong (nu x_1)...(nu x_n)((R Y) p_par S).
% We encode this type via two constructors: one representing the base case without restrictions, and one representing the inductive case
% in which Q cong (nu x)P and Q' cong (nu x)P', where ex_inp_rew P X P' holds inductively.

%% ExInpRew %%
LF ex_inp_rew: proc → names → (names → proc) → type =
  | inp_base: Q cong ((p_in X R) p_par S) → ({y:names} (Q' y) cong ((R y) p_par S))
    → ex_inp_rew Q X Q'
  | inp_ind: Q cong (p_res P) → ({y:names} (Q' y) cong (p_res (P' y)))
    → ({w:names} ex_inp_rew (P w) X \y.(P' y w)) → ex_inp_rew Q X Q'
%% End %%
;


% We prove the lemma which states that processes involved in input transitions (bstep Q (b_in X) Q')
% can be rewritten (up to congruence) in a canonic way (ex_inp_rew Q X Q').
% Before, we need three additional lemmas which prove the result in particular cases.

%% BsInRewPar1 %%
rec bs_in_rew_par1: (g:ctx) {R:[g ⊢ proc]} [g ⊢ ex_inp_rew Q X \y.Q'[..,y]]
  → [g ⊢ ex_inp_rew (Q p_par R) X \y.(Q'[..,y] p_par R[..])] = ...
%% End %%
/ total d (bs_in_rew_par1 _ _ _ _ _ d) /
mlam R ⇒ fn d ⇒ case d of
   | [g ⊢ inp_base C1 \y.C2[..,y]] ⇒ [g ⊢ inp_base (c_trans (c_par C1)
     (c_sym par_assoc)) \y.(c_trans (c_par C2[..,y]) (c_sym par_assoc))]
   | [g ⊢ inp_ind C1 (\y.C2[..,y]) (\w.D1[..,w])] ⇒
     let [g,w:names ⊢ D2[..,w]] = bs_in_rew_par1 [g,w:names ⊢ R[..]]
     [g,w:names ⊢ D1[..,w]] in [g ⊢ inp_ind (c_trans (c_par C1)
     sc_ext_par) (\y.(c_trans (c_par C2[..,y]) sc_ext_par)) (\w.D2[..,w])]
;

rec bs_in_rew_par2:(g:ctx){R:[g ⊢ proc]}[g ⊢ ex_inp_rew Q X \y.Q'[..,y]] → [g ⊢ ex_inp_rew (R p_par Q) X \y.(R[..] p_par Q'[..,y])] = ?
;
%{mlam R ⇒ fn d ⇒ case d of
  | [g ⊢ inp_base C1 \y.C2[..,y]] ⇒ [g ⊢ inp_base (c_trans par_comm (c_trans (c_par C1) (c_sym par_assoc))) \y.(c_trans par_comm (c_trans (c_par C2[..,y]) (c_sym par_assoc)))]
  | [g ⊢ inp_ind C1 (\y.C2[..,y]) (\w.D1[..,w])] ⇒ let [g,w:names ⊢ D2[..,w]] = bs_in_rew_par2 [g,w:names ⊢ R[..]] [g,w:names ⊢ D1[..,w]]
    in [g ⊢ inp_ind (c_trans par_comm (c_trans (c_par C1) (c_trans sc_ext_par (c_res \w.par_comm)))) (\y.(c_trans par_comm (c_trans (c_par C2[..,y]) (c_trans sc_ext_par (c_res \w.par_comm))))) (\w.D2[..,w])]
;}%

rec bs_in_rew_res: (g:ctx) [g,z:names ⊢ ex_inp_rew Q[..,z] X[..] \y.Q'[..,z,y]] → [g ⊢ ex_inp_rew (p_res \z.Q[..,z]) X \y.(p_res \z.Q'[..,z,y])] = ?
;
%{fn d ⇒ case d of
  | [g,z:names ⊢ inp_base C1[..,z] \y.C2[..,z,y]] ⇒ [g ⊢ inp_ind (c_res \z.C1[..,z]) (\y.(c_res \z.C2[..,z,y])) (\z.(inp_base c_ref \y.c_ref))]
  | [g,z:names ⊢ inp_ind C1[..,z] (\y.C2[..,z,y]) (\w.D1[..,z,w])] ⇒ let [g,z:names ⊢ D2[..,z]] = bs_in_rew_res [g,z:names,w:names ⊢
    D1[..,z,w]] in [g ⊢ inp_ind (c_res \z.C1[..,z]) (\y.(c_res \z.C2[..,z,y])) (\z.D2[..,z])]
;}%

%% BsInRew %%
rec bs_in_rew: (g:ctx) [g ⊢ bstep Q (b_in X) \y.Q'[..,y]]
  → [g ⊢ ex_inp_rew Q X \y.Q'[..,y]] =
/ total b (bs_in_rew _ _ _ _ b) /
fn b ⇒ case b of
   | [g ⊢ bs_in] ⇒ [g ⊢ inp_base (c_sym par_unit) \y.(c_sym par_unit)]
   | [g ⊢ bs_par1 B1]:[g ⊢ bstep (P p_par R) (b_in X) (\y.(P' p_par (R[..])))] ⇒
     let [g ⊢ D1] = bs_in_rew [g ⊢ B1] in
     let [g ⊢ D2] = bs_in_rew_par1 [g ⊢ R] [g ⊢ D1] in [g ⊢ D2]
   | [g ⊢ bs_par2 B2]:[g ⊢ bstep (R p_par P) (b_in X) (\y.((R[..]) p_par P'))] ⇒
     let [g ⊢ D1] = bs_in_rew [g ⊢ B2] in 
     let [g ⊢ D2] = bs_in_rew_par2 [g ⊢ R] [g ⊢ D1] in [g ⊢ D2]
   | [g ⊢ bs_res \y.B[..,y]] ⇒
     let [g,y:names ⊢ D1[..,y]] = bs_in_rew [g,y:names ⊢ B[..,y]] in
     let [g ⊢ D2] = bs_in_rew_res [g,y:names ⊢ D1[..,y]] in [g ⊢ D2]
%% End %%
;
%{
/ total q (bs_in_rew _ q _ _ _) /
mlam Q ⇒ fn b ⇒ case [_ ⊢ Q] of
   | [g ⊢ p_zero] ⇒ impossible b
   | [g ⊢ p_in W \z.P[..,z]] ⇒ let [g ⊢ bs_in] = b in
     [g ⊢ inp_base (c_sym par_unit) \y.(c_sym par_unit)]
   | [g ⊢ p_out W Z P] ⇒ impossible b
   | [g ⊢ Q1 p_par Q2] ⇒
     (case b of
        | [g ⊢ bs_par1 B1] ⇒
          let [g ⊢ D1] = bs_in_rew [g ⊢ Q1] [g ⊢ B1] in
          let [g ⊢ D2] = bs_in_rew_par1 [g ⊢ Q2] [g ⊢ D1] in [g ⊢ D2]
        | [g ⊢ bs_par2 B2] ⇒
          let [g ⊢ D1] = bs_in_rew [g ⊢ Q2] [g ⊢ B2] in
          let [g ⊢ D2] = bs_in_rew_par2 [g ⊢ Q1] [g ⊢ D1] in [g ⊢ D2]
      )
   | [g ⊢ p_res \z.P[..,z]] ⇒ let [g ⊢ bs_res \z.B[..,z]] = b in
     let [g,z:names ⊢ D1[..,z]] = bs_in_rew [g,z:names ⊢ P[..,z]]
     [g,z:names ⊢ B[..,z]] in
     let [g ⊢ D2] = bs_in_rew_res [g,z:names ⊢ D1[..,z]] in [g ⊢ D2]
}%
\end{lstlisting}
The other two cases of the main proof follow the same pattern and require similar auxiliary
lemmas.

We are now ready to discuss the proof of 
the main result of this section, namely that $P \xrightarrow{\tau} Q$
implies $P \rightarrow Q$ (Theorem~\ref{thm1}):

\begin{lstlisting}[linerange={FstepImplRed-End}]
% In order to prove that a transition of a process P into Q through a tau action (fstep P f_tau Q) implies reduction of P into Q (P red Q),
% we need four additional lemmas which prove the result in particular cases.
% Each of this lemmas is structured in the same way.


% First lemma - case of the fs_com1 rule:
% If ex_fout_rew P1 X Y Q1 and ex_inp_rew P2 X Q2, then (P1 p_par P2) red (Q1 p_par (Q2 Y)).

% In order to prove this lemma, we analyze the structure of both of the types ex_inp_rew and ex_fout_rew that we have as hypotheses.
% When we consider the inductive case of each type, we need to apply structural induction to proceed:
% so we need a lexicographic induction overall.
% To encode this in Beluga, we need to split the proof in two parts:
% A first lemma (fs_com1_impl_red_base), in which we consider the base case of the first type and proceed by induction on the second type;
% a second lemma (fs_com1_impl_red), in which we consider the inductive case of the first type and proceed by induction on that first type.

%% FsCom1ImplRedBase %%
rec fs_com1_impl_red_base: (g:ctx) [g ⊢ P2 cong ((p_in X \x.R[..,x]) p_par S)]
   → [g,w:names ⊢ Q2[..,w] cong (R[..,w] p_par S[..])]
   → [g ⊢ ex_fout_rew P1 X Y Q1]
   → [g ⊢ (P1 p_par P2) red (Q1 p_par Q2[..,Y])] = ...
%% End %%
/ total d1 (fs_com1_impl_red_base _ _ _ _ _ _ _ _ _ _ _ d1) /
fn c3 ⇒ fn c4 ⇒ fn d1 ⇒ case d1 of
   | [g ⊢ fout_base C1 C2] ⇒
     let [g ⊢ C3] = c3 in
     let [g,w:names ⊢ C4[..,w]] = c4 in
     [g ⊢ r_str (c_trans (c_par C1) (c_trans par_comm (c_trans (c_par C3)
     par_comm))) (r_str par_assoc (r_par (r_str (c_trans (c_par par_comm)
     (c_trans (c_sym par_assoc) par_comm)) (r_par r_com) (c_trans par_comm
     (c_trans par_assoc (c_par par_comm))))) (c_sym par_assoc)) (c_trans
     (c_par (c_sym C2)) (c_trans par_comm (c_trans (c_par
     (c_sym C4[..,_])) par_comm)))]    
   | [g ⊢ fout_ind C1 C2 \w.D1[..,w]] ⇒
     let [g ⊢ C3] = c3 in
     let [g,y:names ⊢ C4[..,y]] = c4 in
     let [g,w:names ⊢ R1[..,w]] = fs_com1_impl_red_base
     [g,w:names ⊢ C3[..]] [g,w:names,y:names ⊢ C4[..,y]]
     [g,w:names ⊢ D1[..,w]] in
     [g ⊢ r_str (c_trans (c_par C1) sc_ext_par) (r_res \w.R1[..,w])
     (c_trans (c_sym sc_ext_par) (c_par (c_sym C2)))]
;

%% FsCom1ImplRed %%
rec fs_com1_impl_red: (g:ctx) [g ⊢ ex_fout_rew P1 X Y Q1]
  → [g ⊢ ex_inp_rew P2 X \x.Q2[..,x]]
  → [g ⊢ (P1 p_par P2) red (Q1 p_par Q2[..,Y])] = ...
%% End %%
/ total d2 (fs_com1_impl_red _ _ _ _ _ _ _ _ d2) /
fn d1 ⇒ fn d2 ⇒ case d2 of
   | [g ⊢ inp_base C3 \y.C4[..,y]] ⇒
     let [g ⊢ R] = fs_com1_impl_red_base [g ⊢ C3]
     [g,y:names ⊢ C4[..,y]] d1 in [g ⊢ R]
   | [g ⊢ inp_ind C3 (\y.C4[..,y]) (\w.D2[..,w])] ⇒
     let [g ⊢ D1] = d1 in
     let [g,w:names ⊢ R1[..,w]] = fs_com1_impl_red [g,w:names ⊢ D1[..]]
     [g,w:names ⊢ D2[..,w]] in
     [g ⊢ r_str (c_trans par_comm (c_trans (c_par C3) (c_trans sc_ext_par
     (c_res \w.par_comm)))) (r_res \w.R1[..,w]) (c_trans (c_res
     \w.par_comm) (c_trans (c_sym sc_ext_par) (c_trans (c_par
     (c_sym C4[..,_])) par_comm)))]
;


% Second lemma - case of the fs_com2 rule:
% If ex_inp_rew P1 X Q1 and ex_fout_rew P2 X Y Q2,
% then (P1 p_par P2) red ((Q1 Y) p_par Q2).
rec fs_com2_impl_red_base: (g:ctx) [g ⊢ P1 cong ((p_in X \x.R[..,x]) p_par S)] → [g,w:names ⊢ Q1[..,w] cong (R[..,w] p_par S[..])] → [g ⊢ ex_fout_rew P2 X Y Q2] → [g ⊢ (P1 p_par P2) red (Q1[..,Y] p_par Q2)] =
/ total d2 (fs_com2_impl_red_base _ _ _ _ _ _ _ _ _ _ _ d2) /
fn c1 ⇒ fn c2 ⇒ fn d2 ⇒ case d2 of
  | [g ⊢ fout_base C3 C4] ⇒ let [g ⊢ C1] = c1 in
    let [g,w:names ⊢ C2[..,w]] = c2 in
    [g ⊢ r_str (c_trans (c_par C1) (c_trans par_comm (c_trans (c_par C3) par_comm))) (r_str par_assoc (r_par (r_str (c_trans (c_par par_comm) (c_trans (c_sym par_assoc) par_comm)) (r_par (r_str par_comm r_com par_comm)) (c_trans par_comm (c_trans par_assoc (c_par par_comm))))) (c_sym par_assoc)) (c_trans (c_par (c_sym C2[..,_])) (c_trans par_comm (c_trans (c_par (c_sym C4)) par_comm)))]
  | [g ⊢ fout_ind C3 C4 \w.D2[..,w]] ⇒ let [g ⊢ C1] = c1 in
    let [g,y:names ⊢ C2[..,y]] = c2 in
    let [g,w:names ⊢ R1[..,w]] = fs_com2_impl_red_base [g,w:names ⊢ C1[..]] [g,w:names,y:names ⊢ C2[..,y]] [g,w:names ⊢ D2[..,w]] in [g ⊢ r_str (c_trans par_comm (c_trans (c_par C3) (c_trans sc_ext_par (c_res \w.par_comm)))) (r_res \w.R1[..,w]) (c_trans (c_res \w.par_comm) (c_trans (c_sym sc_ext_par) (c_trans (c_par (c_sym C4)) par_comm)))]
;

rec fs_com2_impl_red: (g:ctx) [g ⊢ ex_inp_rew P1 X \x.Q1[..,x]] → [g ⊢ ex_fout_rew P2 X Y Q2] → [g ⊢ (P1 p_par P2) red (Q1[..,Y] p_par Q2)] =
/ total d1 (fs_com2_impl_red _ _ _ _ _ _ _ d1 _) /	
fn d1 ⇒ fn d2 ⇒ case d1 of
  | [g ⊢ inp_base C1 \y.C2[..,y]] ⇒ let [g ⊢ R] = fs_com2_impl_red_base [g ⊢ C1] [g,y:names ⊢ C2[..,y]] d2 in [g ⊢ R]
  | [g ⊢ inp_ind C1 (\y.C2[..,y]) (\w.D1[..,w])] ⇒ let [g ⊢ D2] = d2 in
    let [g,w:names ⊢ R1[..,w]] = fs_com2_impl_red [g,w:names ⊢ D1[..,w]] [g,w:names ⊢ D2[..]] in
    [g ⊢ r_str (c_trans (c_par C1) sc_ext_par) (r_res \w.R1[..,w]) (c_trans (c_sym sc_ext_par) (c_par (c_sym C2[..,_])))]			   
;


% Third lemma - case of the fs_close1 rule:
% If ex_bout_rew P1 X Q1 and ex_inp_rew P2 X Q2,
% then (P1 p_par P2) red (nu z) ((Q1 z) p_par (Q2 z)).
rec fs_close1_impl_red_base: (g:ctx) [g ⊢ P2 cong ((p_in X \x.R[..,x]) p_par S)] → [g,w:names ⊢ Q2[..,w] cong (R[..,w] p_par S[..])] → [g ⊢ ex_bout_rew P1 X \x.Q1[..,x]] → [g ⊢ (P1 p_par P2) red (p_res \x.(Q1[..,x] p_par Q2[..,x]))] =
/ total d1 (fs_close1_impl_red_base _ _ _ _ _ _ _ _ _ _ d1) /
fn c3 ⇒ fn c4 ⇒ fn d1 ⇒ case d1 of
  | [g ⊢ bout_base C1 \y.C2[..,y]] ⇒ let [g ⊢ C3] = c3 in
    let [g,w:names ⊢ C4[..,w]] = c4 in
    [g ⊢ r_str (c_trans (c_par C1) sc_ext_par) (r_res \z.(r_str (c_trans par_comm (c_trans (c_par C3[..]) par_comm)) (r_str par_assoc (r_par (r_str (c_trans (c_par par_comm) (c_trans (c_sym par_assoc) par_comm)) (r_par r_com) (c_trans par_comm (c_trans par_assoc (c_par par_comm))))) ((c_sym par_assoc))) (c_trans (c_par (c_sym C2[..,_])) (c_trans par_comm (c_trans (c_par (c_sym C4[..,_])) par_comm))))) (c_ref)]
  | [g ⊢ bout_ind C1 (\y.C2[..,y]) (\w.D1[..,w])] ⇒ let [g ⊢ C3] = c3 in
    let [g,w:names ⊢ C4[..,w]] = c4 in
    let [g,w:names ⊢ R1[..,w]] = fs_close1_impl_red_base [g,w:names ⊢ C3[..]] [g,w:names,y:names ⊢ C4[..,y]] [g,w:names ⊢ D1[..,w]] in
    [g ⊢ r_str (c_trans (c_par C1) sc_ext_par) (r_res \w.R1[..,w]) (c_trans sc_ext_res (c_trans (c_res \z.(c_sym sc_ext_par)) (c_res \z.(c_par (c_sym C2[..,_])))))]
;

rec fs_close1_impl_red: (g:ctx) [g ⊢ ex_bout_rew P1 X \x.Q1[..,x]] → [g ⊢ ex_inp_rew P2 X \x.Q2[..,x]] → [g ⊢ (P1 p_par P2) red (p_res \x.(Q1[..,x] p_par Q2[..,x]))] =
/ total d2 (fs_close1_impl_red _ _ _ _ _ _ _ d2) /
fn d1 ⇒ fn d2 ⇒ case d2 of
  | [g ⊢ inp_base C3 \y.C4[..,y]] ⇒ let [g ⊢ R] = fs_close1_impl_red_base [g ⊢ C3] [g,y:names ⊢ C4[..,y]] d1 in [g ⊢ R]
  | [g ⊢ inp_ind C3 (\y.C4[..,y]) (\w.D2[..,w])] ⇒ let [g ⊢ D1] = d1 in
    let [g,w:names ⊢ R1[..,w]] = fs_close1_impl_red [g,w:names ⊢ D1[..]] [g,w:names ⊢ D2[..,w]] in
    [g ⊢ r_str (c_trans par_comm (c_trans (c_par C3) (c_trans sc_ext_par (c_res \w.par_comm)))) (r_res \w.R1[..,w]) (c_trans sc_ext_res (c_trans (c_res \z.(c_res \w.par_comm)) (c_trans (c_res \z.(c_sym sc_ext_par)) (c_trans (c_res \z.(c_par (c_sym C4[..,_]))) (c_res \z.par_comm)))))]
;


% Fourth lemma - case of the fs_close2 rule:
% If ex_inp_rew P1 X Q1 and ex_bout_rew P2 X Q2,
% then (P1 p_par P2) red (nu z) ((Q1 z) p_par (Q2 z)).
rec fs_close2_impl_red_base: (g:ctx) [g ⊢ P1 cong ((p_in X \x.R[..,x]) p_par S)] → [g,w:names ⊢ Q1[..,w] cong (R[..,w] p_par S[..])] → [g ⊢ ex_bout_rew P2 X \x.Q2[..,x]] → [g ⊢ (P1 p_par P2) red (p_res \x.(Q1[..,x] p_par Q2[..,x]))] =
/ total d2 (fs_close2_impl_red_base _ _ _ _ _ _ _ _ _ _ d2) /
fn c1 ⇒ fn c2 ⇒ fn d2 ⇒ case d2 of
  | [g ⊢ bout_base C3 \y.C4[..,y]] ⇒ let [g ⊢ C1] = c1 in
    let [g,w:names ⊢ C2[..,w]] = c2 in
    [g ⊢ r_str (c_trans par_comm (c_trans (c_par C3) (c_trans sc_ext_par (c_res \z.par_comm)))) (r_res \z.(r_str (c_par C1[..]) (r_str par_assoc (r_par (r_str (c_trans (c_par par_comm) (c_trans (c_sym par_assoc) par_comm)) (r_par (r_str par_comm r_com par_comm)) (c_trans par_comm (c_trans par_assoc (c_par par_comm))))) ((c_sym par_assoc))) (c_trans (c_par (c_sym C2[..,_])) (c_trans par_comm (c_trans (c_par (c_sym C4[..,_])) par_comm))))) (c_ref)]
  | [g ⊢ bout_ind C3 (\y.C4[..,y]) (\w.D2[..,w])] ⇒ let [g ⊢ C1] = c1 in
    let [g,w:names ⊢ C2[..,w]] = c2 in
    let [g,w:names ⊢ R1[..,w]] = fs_close2_impl_red_base [g,w:names ⊢ C1[..]] [g,w:names,y:names ⊢ C2[..,y]] [g,w:names ⊢ D2[..,w]] in
    [g ⊢ r_str (c_trans par_comm (c_trans (c_par C3) (c_trans sc_ext_par (c_res \w.par_comm)))) (r_res \w.R1[..,w]) (c_trans sc_ext_res (c_trans (c_res \z.(c_res \w.par_comm)) (c_trans (c_res \z.(c_sym sc_ext_par)) (c_trans (c_res \z.(c_par (c_sym C4[..,_]))) (c_res \z.par_comm)))))]
;

rec fs_close2_impl_red: (g:ctx) [g ⊢ ex_inp_rew P1 X \x.Q1[..,x]] → [g ⊢ ex_bout_rew P2 X \x.Q2[..,x]] → [g ⊢ (P1 p_par P2) red (p_res \x.(Q1[..,x] p_par Q2[..,x]))] =
/ total d1 (fs_close2_impl_red _ _ _ _ _ _ d1 _) /
fn d1 ⇒ fn d2 ⇒ case d1 of
  | [g ⊢ inp_base C1 \y.C2[..,y]] ⇒ let [g ⊢ R] = fs_close2_impl_red_base [g ⊢ C1] [g,y:names ⊢ C2[..,y]] d2 in [g ⊢ R]
  | [g ⊢ inp_ind C1 (\y.C2[..,y]) (\w.D1[..,w])] ⇒ let [g ⊢ D2] = d2 in
    let [g,w:names ⊢ R1[..,w]] = fs_close2_impl_red [g,w:names ⊢ D1[..,w]] [g,w:names ⊢ D2[..]] in
    [g ⊢ r_str (c_trans (c_par C1) sc_ext_par) (r_res \w.R1[..,w]) (c_trans sc_ext_res (c_trans (c_res \w.(c_sym sc_ext_par)) (c_res \w.(c_par (c_sym C2[..,_])))))]
;



% We now prove the first theorem: a transition of a process P into Q through a tau action (fstep P f_tau Q) implies reduction of P into Q (P red Q).

%% FstepImplRed %%
rec fstep_impl_red: (g:ctx) [g ⊢ fstep P f_tau Q] → [g ⊢ P red Q] = ...
%% End %%
/ total f (fstep_impl_red _ _ _ f) /
fn f ⇒ case f of
   | [g ⊢ fs_par1 F1] ⇒ let [g ⊢ R] = fstep_impl_red [g ⊢ F1] in
                         [g ⊢ r_par R]
   | [g ⊢ fs_par2 F2] ⇒ let [g ⊢ R] = fstep_impl_red [g ⊢ F2] in
                         [g ⊢ r_str par_comm (r_par R) par_comm] 
   | [g ⊢ fs_com1 F1 B1] ⇒
     let [g ⊢ D1] = fs_out_rew [g ⊢ _] [g ⊢ F1] in
     let [g ⊢ D2] = bs_in_rew [g ⊢ _] [g ⊢ B1] in
     let [g ⊢ R] = fs_com1_impl_red [g ⊢ D1] [g ⊢ D2] in [g ⊢ R]
   | [g ⊢ fs_com2 B1 F1] ⇒
     let [g ⊢ D1] = bs_in_rew [g ⊢ _] [g ⊢ B1] in
     let [g ⊢ D2] = fs_out_rew [g ⊢ _] [g ⊢ F1] in
     let [g ⊢ R] = fs_com2_impl_red [g ⊢ D1] [g ⊢ D2] in [g ⊢ R]
   | [g ⊢ fs_res \z.F[..,z]] ⇒
     let [g,z:names ⊢ R[..,z]] = fstep_impl_red [g,z:names ⊢ F[..,z]] in 
     [g ⊢ r_res \z.R[..,z]]
   | [g ⊢ fs_close1 B1 B2] ⇒
     let [g ⊢ D1] = bs_out_rew [g ⊢ _][g ⊢ B1] in
     let [g ⊢ D2] = bs_in_rew [g ⊢ _] [g ⊢ B2] in
     let [g ⊢ R] = fs_close1_impl_red [g ⊢ D1] [g ⊢ D2] in [g ⊢ R]
   | [g ⊢ fs_close2 B1 B2] ⇒
     let [g ⊢ D1] = bs_in_rew [g ⊢ _] [g ⊢ B1] in
     let [g ⊢ D2] = bs_out_rew [g ⊢ _] [g ⊢ B2] in
     let [g ⊢ R] = fs_close2_impl_red [g ⊢ D1] [g ⊢ D2] in [g ⊢ R]
;
\end{lstlisting}

The proof proceeds by induction on the derivation 
of 
{\small \verb|[g |$\vdash$\verb| fstep P f_tau Q]|}. Mirroring the informal proof, in certain subcases it is enough
to apply a recursive call of the function on a structurally smaller
$\tau$-transition and return the desired object of type {\small
  \verb|[g |$\vdash$\verb| P red Q]|}. In the other subcases, we apply
the functions {\small \color{belugablue} \verb|bs_in_rew|}, {\small
  \color{belugablue} \verb|fs_out_rew|} and {\small \color{belugablue}
  \verb|bs_out_rew|} on the given input/output/bound output
transitions, obtaining the corresponding objects of existential type, viz.\ the cited {\small
  \color{belugagreen}\verb|ex_inp_rew|} and its ``siblings''
  {\small
   \color{belugagreen}\verb|ex_fout_rew|} and {\small
   \color{belugagreen}\verb|ex_bout_rew|}.
 Then, we would like to conclude by applying some auxiliary functions
 which unfold these objects and build the desired reduction. Here, we
 face a major hurdle, not so much in writing down the proof terms, but
 in having them checked for \emph{termination}, which is, after all,
 what guarantees that the inductive structure of the proof is
 correct. Let us see one of such lemmas, which emerges in the subcase
 where a {\small \textsc{S-Com-L}} rule is applied:

 \lstinputlisting[linerange={FsCom1ImplRed-End}]{BelugaCode/5-theorem1.bel}
 The proof needs to consider both hypotheses, in order to unfold the
 two existential types -- recall that the telescopes force upon us an
 inductive encoding of those existentials. In other terms,
 verification of the fact that this function is decreasing would need to appeal 
 to a form of \emph{lexicographic} induction.  

 Termination checkers are of course incomplete and adopt strict
 syntactic criteria to enforce it; in particular, Beluga's checker
 currently does not support lexicographic induction. One way out is to
 define the {\small \color{belugablue} \verb|fs_com1_impl_red|}
 function so that it is decreasing on the first argument only, and 
 relies on an auxiliary function addressing its base case that is
 decreasing on the second argument. The signature of such a function
 is:
 \lstinputlisting[linerange={FsCom1ImplRedBase-End}]{BelugaCode/5-theorem1.bel}

Once these lemmas are in place, the proof of the first direction of the Harmony Lemma follows without any further drama.

\subsubsection{Theorem~\ref{thm2}: Reduction Implies $\tau$-Transition}
In the other direction,  having HOAS disposed of most of the technical
lemmas about names and substitutions, the workhorses are the reduction
rewriting Lemma~\ref{lm27} and the congruence-as-bisimilarity
Lemma~\ref{lm26}. Since the former does not introduce new ideas, we discuss it
first.

Given its similarity to Lemmas~\ref{lm11}--\ref{lm13}, Lemma~\ref{lm27}
is implemented with the same strategy: we define an existential type
{\small \color{belugagreen}\verb|ex_red_rew|} that inductively 
 encodes the existence of the telescopes and the two
 congruences stated in the conclusion of the lemma.
\begin{lstlisting}[linerange={ExRedRew-End}]
% Predicate about syntactic equivalence of processes involved in reductions:
% ex_red_rew P Q holds iff there exist R1, R2, S, X, Y, x_1,...,x_n such that P cong (nu x_1)...(nu x_n)(((p_out X Y R1) p_par (p_in X R2)) p_par S) and
% Q cong (nu x_1)...(nu x_n)((R1 p_par (R2 Y)) p_par S).
% We encode this type via two constructors: one representing the base case without restrictions, and one representing the inductive case
% in which P cong (nu x)P' and Q cong (nu x)Q', where ex_red_rew P' Q' holds inductively.

%% ExRedRew %%
LF ex_red_rew: proc → proc → type =
  | red_base: P cong (((p_out X Y R1) p_par (p_in X R2)) p_par S)
    → Q cong ((R1 p_par (R2 Y)) p_par S) → ex_red_rew P Q
  | red_ind: P cong (p_res P') → Q cong (p_res Q')
    → ({w:names} ex_red_rew (P' w) (Q' w)) → ex_red_rew P Q
%% End %%
;

% We prove the lemma which states that processes involved in reductions (P red Q)
% can be rewritten (up to congruence) in a canonic way (ex_red_rew P Q).
% Before, we need three additional lemmas which prove the result in particular cases.

%% RedImplRedRewPar %%
rec red_impl_red_rew_par: (g:ctx) {R:[g ⊢ proc]} [g ⊢ ex_red_rew P Q]
  → [g ⊢ ex_red_rew (P p_par R) (Q p_par R)] = ...
%% End %%
/ total d (red_impl_red_rew_par _ _ _ _ d) /
mlam R ⇒ fn d ⇒ case d of
   | [g ⊢ red_base C1 C2] ⇒
     [g ⊢ red_base (c_trans (c_par C1) (c_sym par_assoc))
     (c_trans (c_par C2) (c_sym par_assoc))]
   | [g ⊢ red_ind C1 C2 \w.D1[..,w]] ⇒
     let [g,w:names ⊢ D2[..,w]] = red_impl_red_rew_par
     [g,w:names ⊢ R[..]] [g,w:names ⊢ D1[..,w]] in
     [g ⊢ red_ind (c_trans (c_par C1) sc_ext_par) 
     (c_trans (c_par C2) sc_ext_par) \w.D2[..,w]]
;

rec red_impl_red_rew_res: (g:ctx) [g, z:names ⊢ ex_red_rew P[..,z] Q[..,z]] → [g ⊢ ex_red_rew (p_res \z.P[..,z]) (p_res \z.Q[..,z])] =
/ total d (red_impl_red_rew_res _ _ _ d)/
fn d ⇒ case d of
  | [g,z:names ⊢ red_base C1[..,z] C2[..,z]] ⇒ [g ⊢ red_ind c_ref c_ref \z.(red_base C1[..,z] C2[..,z])]
  | [g,z:names ⊢ red_ind C1[..,z] C2[..,z] \w.D1[..,z,w]] ⇒ let [g,z:names ⊢ D2[..,z]] = red_impl_red_rew_res [g,z:names,w:names ⊢ D1[..,z,w]] in [g ⊢ red_ind (c_res \z.C1[..,z]) (c_res \z.C2[..,z]) \z.D2[..,z]]
;

rec red_impl_red_rew_str: (g:ctx) [g ⊢ P cong P'] → [g ⊢ ex_red_rew P' Q'] → [g ⊢ Q' cong Q] → [g ⊢ ex_red_rew P Q] =
/ total d (red_impl_red_rew_str _ _ _ _ _ _ d _)/
fn c1 ⇒ fn d ⇒ fn c2 ⇒ case d of
  | [g ⊢ red_base C1' C2'] ⇒ let [g ⊢ C1] = c1 in
			     let [g ⊢ C2] = c2 in [g ⊢ red_base (c_trans C1 C1') (c_trans (c_sym C2) C2')]
  | [g ⊢ red_ind C1' C2' \w.D1[..,w]] ⇒ let [g ⊢ C1] = c1 in
					 let [g ⊢ C2] = c2 in
					 [g ⊢ red_ind (c_trans C1 C1') (c_trans (c_sym C2) C2') \w.D1[..,w]]
;

%% RedImplRedRew %%
rec red_impl_red_rew: (g:ctx) [g ⊢ P red Q] → [g ⊢ ex_red_rew P Q] = ...
%% End %%
/ total r (red_impl_red_rew _ _ _ r) /
fn r ⇒ case r of
  | [g ⊢ r_com] ⇒ [g ⊢ red_base (c_sym par_unit) (c_sym par_unit)]
  | [g ⊢ r_par R1]:[g ⊢ (P p_par R) red (Q p_par R)] ⇒
    let [g ⊢ D1] = red_impl_red_rew [g ⊢ R1] in
    let [g ⊢ D2] = red_impl_red_rew_par [g ⊢ R] [g ⊢ D1] in [g ⊢ D2]
  | [g ⊢ r_res \z.R1[..,z]] ⇒
    let [g,z:names ⊢ D1[..,z]] = red_impl_red_rew
    [g,z:names ⊢ R1[..,z]] in
    let [g ⊢ D2] = red_impl_red_rew_res [g,z:names ⊢ D1[..,z]] in
    [g ⊢ D2]
  | [g ⊢ r_str C1 R1 C2] ⇒
    let [g ⊢ D1] = red_impl_red_rew [g ⊢ R1] in
    let [g ⊢ D2] = red_impl_red_rew_str [g ⊢ C1] [g ⊢ D1] [g ⊢ C2] in
    [g ⊢ D2]
;
\end{lstlisting}
We then implement some  auxiliary
functions to unfold objects of the existential type {\small
  \color{belugagreen}\verb|ex_red_rew|} in specific subcases, for example:
\begin{lstlisting}[linerange={RedImplRedRewPar-End}]
% Predicate about syntactic equivalence of processes involved in reductions:
% ex_red_rew P Q holds iff there exist R1, R2, S, X, Y, x_1,...,x_n such that P cong (nu x_1)...(nu x_n)(((p_out X Y R1) p_par (p_in X R2)) p_par S) and
% Q cong (nu x_1)...(nu x_n)((R1 p_par (R2 Y)) p_par S).
% We encode this type via two constructors: one representing the base case without restrictions, and one representing the inductive case
% in which P cong (nu x)P' and Q cong (nu x)Q', where ex_red_rew P' Q' holds inductively.

%% ExRedRew %%
LF ex_red_rew: proc → proc → type =
  | red_base: P cong (((p_out X Y R1) p_par (p_in X R2)) p_par S)
    → Q cong ((R1 p_par (R2 Y)) p_par S) → ex_red_rew P Q
  | red_ind: P cong (p_res P') → Q cong (p_res Q')
    → ({w:names} ex_red_rew (P' w) (Q' w)) → ex_red_rew P Q
%% End %%
;

% We prove the lemma which states that processes involved in reductions (P red Q)
% can be rewritten (up to congruence) in a canonic way (ex_red_rew P Q).
% Before, we need three additional lemmas which prove the result in particular cases.

%% RedImplRedRewPar %%
rec red_impl_red_rew_par: (g:ctx) {R:[g ⊢ proc]} [g ⊢ ex_red_rew P Q]
  → [g ⊢ ex_red_rew (P p_par R) (Q p_par R)] = ...
%% End %%
/ total d (red_impl_red_rew_par _ _ _ _ d) /
mlam R ⇒ fn d ⇒ case d of
   | [g ⊢ red_base C1 C2] ⇒
     [g ⊢ red_base (c_trans (c_par C1) (c_sym par_assoc))
     (c_trans (c_par C2) (c_sym par_assoc))]
   | [g ⊢ red_ind C1 C2 \w.D1[..,w]] ⇒
     let [g,w:names ⊢ D2[..,w]] = red_impl_red_rew_par
     [g,w:names ⊢ R[..]] [g,w:names ⊢ D1[..,w]] in
     [g ⊢ red_ind (c_trans (c_par C1) sc_ext_par) 
     (c_trans (c_par C2) sc_ext_par) \w.D2[..,w]]
;

rec red_impl_red_rew_res: (g:ctx) [g, z:names ⊢ ex_red_rew P[..,z] Q[..,z]] → [g ⊢ ex_red_rew (p_res \z.P[..,z]) (p_res \z.Q[..,z])] =
/ total d (red_impl_red_rew_res _ _ _ d)/
fn d ⇒ case d of
  | [g,z:names ⊢ red_base C1[..,z] C2[..,z]] ⇒ [g ⊢ red_ind c_ref c_ref \z.(red_base C1[..,z] C2[..,z])]
  | [g,z:names ⊢ red_ind C1[..,z] C2[..,z] \w.D1[..,z,w]] ⇒ let [g,z:names ⊢ D2[..,z]] = red_impl_red_rew_res [g,z:names,w:names ⊢ D1[..,z,w]] in [g ⊢ red_ind (c_res \z.C1[..,z]) (c_res \z.C2[..,z]) \z.D2[..,z]]
;

rec red_impl_red_rew_str: (g:ctx) [g ⊢ P cong P'] → [g ⊢ ex_red_rew P' Q'] → [g ⊢ Q' cong Q] → [g ⊢ ex_red_rew P Q] =
/ total d (red_impl_red_rew_str _ _ _ _ _ _ d _)/
fn c1 ⇒ fn d ⇒ fn c2 ⇒ case d of
  | [g ⊢ red_base C1' C2'] ⇒ let [g ⊢ C1] = c1 in
			     let [g ⊢ C2] = c2 in [g ⊢ red_base (c_trans C1 C1') (c_trans (c_sym C2) C2')]
  | [g ⊢ red_ind C1' C2' \w.D1[..,w]] ⇒ let [g ⊢ C1] = c1 in
					 let [g ⊢ C2] = c2 in
					 [g ⊢ red_ind (c_trans C1 C1') (c_trans (c_sym C2) C2') \w.D1[..,w]]
;

%% RedImplRedRew %%
rec red_impl_red_rew: (g:ctx) [g ⊢ P red Q] → [g ⊢ ex_red_rew P Q] = ...
%% End %%
/ total r (red_impl_red_rew _ _ _ r) /
fn r ⇒ case r of
  | [g ⊢ r_com] ⇒ [g ⊢ red_base (c_sym par_unit) (c_sym par_unit)]
  | [g ⊢ r_par R1]:[g ⊢ (P p_par R) red (Q p_par R)] ⇒
    let [g ⊢ D1] = red_impl_red_rew [g ⊢ R1] in
    let [g ⊢ D2] = red_impl_red_rew_par [g ⊢ R] [g ⊢ D1] in [g ⊢ D2]
  | [g ⊢ r_res \z.R1[..,z]] ⇒
    let [g,z:names ⊢ D1[..,z]] = red_impl_red_rew
    [g,z:names ⊢ R1[..,z]] in
    let [g ⊢ D2] = red_impl_red_rew_res [g,z:names ⊢ D1[..,z]] in
    [g ⊢ D2]
  | [g ⊢ r_str C1 R1 C2] ⇒
    let [g ⊢ D1] = red_impl_red_rew [g ⊢ R1] in
    let [g ⊢ D2] = red_impl_red_rew_str [g ⊢ C1] [g ⊢ D1] [g ⊢ C2] in
    [g ⊢ D2]
;
\end{lstlisting}
Having established those, it is straightforward to prove
\lstinputlisting[linerange={RedImplRedRew-End}]{BelugaCode/7-reduction-rewriting.bel}
by induction on the
structure of the given reduction judgment. 

Moving on to the proof of Lemma~\ref{lm26}, there is a new technicality to address.
We have seen how in a HOAS setting provisos such as ``$x \notin$
{\myfont fn($P$)}'' are realized by $P$ being in the scope of a meta
level abstraction that binds $x$, but \emph{not} actually depending on
$x$. Sometimes (see~\cite{BP23} for other instances), we have to convince our proof environment of this non-dependency, which is basically the content of  Lemma~\ref{lm24}, in words:
 ``given a transition
$P \xrightarrow{\alpha} Q$ where $x$ does not occur free in $P$, then $x$ does
not occur free in $\alpha$ and $Q$''.
Since in Beluga judgments over open terms are encapsulated in the context where they make sense, removing these spurious dependencies amounts to ``strengthening'' such a context, akin to strengthening
lemmas in type theory. The lemma more formally reads as:
\begin{equation}
  \label{eq:stren}
  \mbox{If }\Gamma, x : names\, \vdash P \xrightarrow{\alpha_x} Q_x, \mbox{ then there are } \alpha',Q'
  \mbox{ such that } \alpha_x = \alpha', \,Q_x = Q' \mbox{ and }
  \Gamma\,\vdash P \xrightarrow{\alpha'} Q'
\end{equation}

Not only Beluga does not have existentials (nor conjunctions), but LF also has no built-in
equality notion. However, this is easy to simulate via pattern unification over canonical forms, by
defining three type families encoding equality of processes, free
actions and bound actions respectively. Process equality is defined as:

\lstinputlisting[linerange={Eqp-End}]{BelugaCode/6-stepcong-lemma.bel}

Since we are now dealing with a property about a LF contextual object
(the initial transition), the statement in (\ref{eq:stren}) has to be
encoded at the computation level as an \emph{inductive}
type~\cite{DBLP:conf/cade/PientkaC15}.
We list this encoding, omitting the definition of its counterpart for
bound transitions {\small \color{belugagreen}\verb|ex_str_bstep|}.
\lstinputlisting[linerange={ExStrFstep-End}]{BelugaCode/6-stepcong-lemma.bel}
Note how non-occurrence is realized using \emph{weakening} substitutions, i.e.\ 
{\small \verb|P[..]|} signals that the process {\small \verb|P|} depends only on the variables mentioned
in {\small \verb|g|}, excluding {\small \verb|x|}.

The strengthening lemma is implemented through the two following mutually recursive functions:
\lstinputlisting[linerange={Strengthening-End}]{BelugaCode/6-stepcong-lemma.bel}
The proof follows by a long but straightforward induction on the structure of the given transition.

\smallskip To state Lemma~\ref{lm26}, we again need to code the
existential in its conclusion; however, since the statement involves
transitions through a generic action $\alpha$, we actually require two
new type families: one for free transitions and one for bound
transitions.
\lstinputlisting[linerange={ExStepcong-End}]{BelugaCode/6-stepcong-lemma.bel}
%
The reader may wonder why the process {\small \verb|P|} is mentioned
in the {\small \verb|fsc|} and {\small \verb|bsc|} constructors, as it
does not play any role: indeed, it is a trick to please Beluga's coverage checker when this definition is
unfolded in the rest of the development.

Lemma~\ref{lm26} is encoded through the definition of four mutually
recursive functions: as the statement involves transitions via a
generic action $\alpha$, the first two  prove the
result for free transitions, while the last two  demonstrate
the result for bound transitions; moreover, since the proof is carried
out by concurrently establishing two symmetrical assertions, the odd
functions prove the left-to-right statement, while the even ones
demonstrate the right-to-left statement. We present the signature of
the first function {\small \color{belugablue}
  \verb|cong_fstepleft_impl_stepright|}, which proves the
left-to-right assertion for free transitions:
\lstinputlisting[linerange={CongFstepleftImplFstepright-End}]{BelugaCode/6-stepcong-lemma.bel}
Mirroring the informal proof, this lemma is proved by a long induction
on the structure of the given congruence; in most of the subcases,
case analysis of the given transition is performed as well.

A final ingredient for the proof of Theorem~\ref{thm2} is the auxiliary lemma:
\begin{lstlisting}[linerange={RedRewImplFstepcong-End}]
% We now prove that, given two processes P and Q involved in a reduction (P red Q), then there exists a process Q'
% such that P steps into Q' through tau and Q is congruent to Q' (overall, ex_fstepcong P P f_tau Q holds).
% We obtain the result by rewriting the processes P and Q up to congruence in a canonic way (ex_red_rew P Q)
% and stating a lemma which unfolds the type ex_red_rew and reaches the conclusion.

%% RedRewImplFstepcong %%
rec red_rew_impl_fstepcong: (g:ctx) [g ⊢ ex_red_rew P Q]
  → [g ⊢ ex_fstepcong P P f_tau Q] = ...
%% End %%
/ total d (red_rew_impl_fstepcong _ _ _ d) /
fn d ⇒ case d of
   | [g ⊢ red_base C1 C2] ⇒
     let [g ⊢ fsc F C3] = cong_fstepright_impl_fstepleft
     [g ⊢ C1] [g ⊢ fs_par1 (fs_com1 fs_out bs_in)] in
     [g ⊢ fsc F (c_trans C2 C3)]
   | [g ⊢ red_ind C1 C2 \w.D1[..,w]] ⇒
     let [g,w:names ⊢ fsc F1[..,w] C3[..,w]] =
     red_rew_impl_fstepcong [g,w:names ⊢ D1[..,w]] in
     let [g ⊢ fsc F2 C4] = cong_fstepright_impl_fstepleft
     [g ⊢ C1] [g ⊢ fs_res \w.F1[..,w]] in
     [g ⊢ fsc F2 (c_trans C2 (c_trans (c_res \w.C3[..,w]) C4))]
;

%% RedImplFstepcong %%
rec red_impl_fstepcong: (g:ctx) [g ⊢ P red Q] → [g ⊢ ex_fstepcong P P f_tau Q] =
/ total r (red_impl_fstepcong _ _ _ r) /
fn r ⇒ let [g ⊢ D1] = red_impl_red_rew r in
        let [g ⊢ D2] = red_rew_impl_fstepcong [g ⊢ D1] in [g ⊢ D2]
%% End %%
;
\end{lstlisting}
The proof proceeds by induction on the structure of the given object of type {\small \verb|[g |$\vdash$\verb| ex_red_rew P Q]|}; in both the base case and the inductive case, a key factor consists in the application of the function {\small \color{belugablue} \verb|cong_fstepright_impl_fstepleft|}.

We are ready to prove Theorem~\ref{thm2}:
\begin{lstlisting}[linerange={RedImplFstepcong-End}]
% We now prove that, given two processes P and Q involved in a reduction (P red Q), then there exists a process Q'
% such that P steps into Q' through tau and Q is congruent to Q' (overall, ex_fstepcong P P f_tau Q holds).
% We obtain the result by rewriting the processes P and Q up to congruence in a canonic way (ex_red_rew P Q)
% and stating a lemma which unfolds the type ex_red_rew and reaches the conclusion.

%% RedRewImplFstepcong %%
rec red_rew_impl_fstepcong: (g:ctx) [g ⊢ ex_red_rew P Q]
  → [g ⊢ ex_fstepcong P P f_tau Q] = ...
%% End %%
/ total d (red_rew_impl_fstepcong _ _ _ d) /
fn d ⇒ case d of
   | [g ⊢ red_base C1 C2] ⇒
     let [g ⊢ fsc F C3] = cong_fstepright_impl_fstepleft
     [g ⊢ C1] [g ⊢ fs_par1 (fs_com1 fs_out bs_in)] in
     [g ⊢ fsc F (c_trans C2 C3)]
   | [g ⊢ red_ind C1 C2 \w.D1[..,w]] ⇒
     let [g,w:names ⊢ fsc F1[..,w] C3[..,w]] =
     red_rew_impl_fstepcong [g,w:names ⊢ D1[..,w]] in
     let [g ⊢ fsc F2 C4] = cong_fstepright_impl_fstepleft
     [g ⊢ C1] [g ⊢ fs_res \w.F1[..,w]] in
     [g ⊢ fsc F2 (c_trans C2 (c_trans (c_res \w.C3[..,w]) C4))]
;

%% RedImplFstepcong %%
rec red_impl_fstepcong: (g:ctx) [g ⊢ P red Q] → [g ⊢ ex_fstepcong P P f_tau Q] =
/ total r (red_impl_fstepcong _ _ _ r) /
fn r ⇒ let [g ⊢ D1] = red_impl_red_rew r in
        let [g ⊢ D2] = red_rew_impl_fstepcong [g ⊢ D1] in [g ⊢ D2]
%% End %%
;
\end{lstlisting}
%
\sloppypar
Given {\small \verb|r|} representing the reduction $P \rightarrow
Q$, we apply the function {\small \color{belugablue}
  \verb|red_impl_red_rew|} to it  returning an object
{\small \verb|D1|} of type {\small \verb|[g |$\vdash$\verb| ex_red_rew P Q]|} that encodes the following
congruences:  
\mbox{$P \equiv (\nu w_1) \cdots (\nu w_n) ((\bar{x}y.R_1 \mid x(z).R_2) \mid
S)$} and
$Q \equiv (\nu w_1) \cdots (\nu w_n) ((R_1 \mid R_2 \{ y/z \}) \mid S)$, for some $R_1$, $R_2$, $S$ and $w_1, \ldots, w_n$.
 Finally, we invoke the auxiliary function {\small \color{belugablue}
  \verb|red_rew_impl_fstepcong|}, which unfolds the argument {\small
  \verb|D1|} and returns the desired object of type
{\small \verb|[g |$\vdash$\verb| ex_fstepcong P P f_tau Q]|}.


\section{Evaluation and Conclusions}\label{sec:conc}

The reader may wonder ``Is that it?'' We sympathize with the feeling: what is remarkable in
this formalization is how (mostly) uneventful it has been. Once we had
settled on using (weak) HOAS and a specialized proof environment such as Beluga --- which, given our lineage, was not
much of a stretch --- the Honsell/Miller encoding of the labelled
transition system removed all issues related to scope extrusion and
Beluga's remarkable conciseness did the rest, turning $30$ pages of
\LaTeX\ proofs, which still skip many steps, into some $700$ lines of
proof terms.

Remarkably, the structure of the
formal proof closely mirrors the informal one:
having eliminated the 7 technical lemmas thanks to the HOAS encoding, both
proofs share the same 6 lemmas, proved in the same fashion.
Some parts of the formal proof are covered by 22 additional lemmas
which deal with the unfolding of the existential types,\footnote{Given how widespread
existential (and conjunctive) statements are in this development, it would be helpful if
Beluga could provide some syntactic sugar, similarly to Agda, and a
way to unfold those definitions automatically.} while another 4
lemmas result from the mutual recursion induced by the encoding.

In our biased opinion, this uneventfulness does not trivialize the
accomplishment: we have provided a compact and elegant solution of a
benchmark problem, which, after all, is supposed to be challenging:
the fact that this has not been a heroic feat is a testament to the
merits of the HOAS encoding and to the long line of research that has
made meta-level reasoning over HOAS specifications possible. It also
suggests that, once we put on the HOAS ``spectacles'', the binders of
the $\pi$-calculus are not that different from those of the
$\lambda$-calculus, in the sense that the meta-level binder will
gladly model scope extrusion, under the right encoding.

Beluga has shown to be a reliable system: we did stress the
termination checker, with a heavy use of mutually defined recursive
functions. We managed to get around the current lack of support for
lexicographic induction; 
our technique could have broader applicability, for instance  in
verifying totality results such as the admissibility of cut
elimination that rely on nested induction for their termination
proofs~\cite{Pfenning00}. 
%
We also established coverage, again getting around the minor glitch that we have
mentioned above in the proof of the right-to-left direction of the
Harmony Lemma.

Since the benchmark is amenable to a weak HOAS encoding, this begs the
question of why not pursue its solution in a general proof assistant
such as Coq.  While weak HOAS is indeed consistent with monotonic
inductive types, it is well known that the full dependent function
space of a theory such as the Calculus of Constructions is
incompatible with the intensional quantification featured by LF-like
type theories. Workarounds exist: the most successful one is
the ``Theory of Contexts''~\cite{DBLP:journals/tcs/HonsellMS01}, where
additional predicates concerning freshness and non-occurrence of names are
added to specifications such as of the LTS\@. Further, ToC assumes
some axioms regulating the properties of names and abstraction over
names (i.e.\ ``contexts'') in order to reify what LF-like frameworks
provide natively. To be fair, it is unclear to us which role ToC would
play in a Coq solution of CCFB.2, but for the rest of the Concurrent
Benchmark, the outcome is not pretty (believe us, we tried).  Of
course, there is no obstacle in abandoning HOAS for concrete encodings
and we are looking forward to comparing such solutions to ours.

\smallskip Although CCFB.2 does not ask for it, we conjecture that it
would be easy, albeit tedious, to extend our solution to account for
other features of the $\pi$-calculus, namely sums, replication and
match. Mismatch, which is handled
in~\cite{DBLP:journals/tcs/HonsellMS01}, is rather problematic in
HOAS, since the systems that support it have no native notion of negation.

The adoption of Beluga as a proof environment for this formalization
is motivated by our endeavor (together with Pientka's group) to give
an overall HOAS solution to all the challenges listed in CCFB, which
include type safety for (linear) session types and turning strong
barbed bisimilarity into a congruence. For this, Beluga is a strong
candidate: in fact the type safety challenge is already in the bag,
thanks to the techniques developed in~\cite{BP23}. The coinduction part
is more challenging, but we have a good track record in a similar
benchmark~\cite{MomiglianoPT19}.  Solving the rest of CCFB in Beluga
may also shed some light on the role of the $\nabla$
quantifier~\cite{DBLP:journals/tocl/MillerT05} as a meta-reasoning
tool,  compared to  Beluga's use of contextual LF as a specification
language.
\bibliographystyle{eptcs}
\bibliography{References}

\appendix

\section{Appendix: Late vs Early Transitions}
\label{app:equiv}

An alternative to the late semantics presented in the paper is the \emph{early} semantics. As its name suggests, it is characterized by the fact that substitutions of names received in interaction are performed as soon as possible, namely during the execution of the {\small \textsc{S-In}} rule. 
The syntax of actions is the same as in the late semantics; however, the name $y$ occurring in an input action $x(y)$ is considered to be free. As for transitions, now denoted as $\xrightharpoondown{(-)}$, the rules {\small \textsc{S-In}}, the two {\small \textsc{S-Com}} rules and the two {\small \textsc{S-Close}} are replaced by those in Fig.~\ref{fig:early} (``right'' rules are omitted for brevity). Note how the {\small \textsc{S-In}} rule now exhibits the substitution of the input name; conversely, in the {\small \textsc{S-Com}} rules, both of the names $x$ and $y$ occurring in the actions of the given transitions must coincide and no substitution takes place in the conclusion. 
\begin{figure}[th]
\centering
\fbox{\vbox {\advance \hsize by -2\fboxsep \advance \hsize by -2\fboxrule \linewidth\hsize
\begin{mathpar}
  \inferrule[$\mkern 61mu$ S-In] { } {x(z).P \, \xrightharpoondown{x(y)} \, P\{ y/z \}} 
  \and \inferrule[$\mkern 51mu$ S-Com-L] {P \xrightharpoondown{\bar{x}y} P' \\ Q \xrightharpoondown{x(y)} Q'} {P \mid Q \, \xrightharpoondown{\tau} \, P' \mid Q'} 
   \and \inferrule[$\mkern 92mu$ S-Close-L] {P \xrightharpoondown{\bar{x}(z)} P' \quad Q \xrightharpoondown{x(z)} Q' \quad z \notin$ {\myfont fn}$(Q)} {P \mid Q \ \xrightharpoondown{\tau} \ (\nu z) (P' \mid Q')}
\end{mathpar}
}}
\caption{Early transition semantics rules.}
\label{fig:early}
\end{figure}

In order to prove the equivalence of the two semantics, we follow
Parrow's approach in~\cite{DBLP:books/el/01/Parrow01}: namely,
our objective is to demonstrate that 
the two semantics allow to infer the same $\tau$-transitions.
Both directions are achieved
by induction on the depth of the inference of the given transitions
and rely on some additional lemmas, which state
a correspondence between input/output transitions of the two semantics as well.
The only non-trivial correspondence lies
between input (in the late semantics) and free input (in the early
semantics), which is defined as follows:
\begin{equation}
  \label{eq:star}
P \xrightharpoondown{x(y)} Q\  \mbox{
 iff there are $Q'$ and $w$ such that } 
P \xrightarrow{x(w)} Q'  \mbox{ and } Q=Q' \{ y/w \}. 
\end{equation}

We begin the Beluga formalization by encoding the two semantics in the same environment (Fig.~\ref{fig:newlts}). The type {\small \color{belugagreen}\verb|f_act|} presents a new constructor {\small \verb|f_in|} for free input actions in the early semantics; as for transitions, the constructors expressing identical rules in the two semantics are omitted for brevity. We observe that the {\small \verb|ebs_in|} constructor representing bound input transitions in the early semantics cannot be eliminated, since bound input transitions are needed as premises in the rules introduced by the {\small \verb|efs_close|} constructors. For consistency of notation, the types {\small \color{belugagreen}\verb|fstep|} and {\small \color{belugagreen}\verb|bstep|} for late transitions have been renamed as {\small \color{belugagreen}\verb|late_fstep|} and {\small \color{belugagreen}\verb|late_bstep|}.
\begin{figure}[ht]
\lstinputlisting[style=belugatwo,linerange={Act-End,EarlyF-End,EarlyB-End}]{BelugaCode/early-late-equivalence.bel}
\vspace{-0.5\baselineskip}
\caption{Encoding of actions and early transition.}
\label{fig:newlts}
\end{figure}

The next ingredient for the formalization of the semantics equivalence is the definition of the type family {\small \color{belugagreen}\verb|ex_latebs|}, which encodes the existence of a late transition such as in the correspondence (\ref{eq:star}):
\lstinputlisting[linerange={ExLatebs-End}]{BelugaCode/early-late-equivalence.bel}

We then list the signature of the functions {\small \color{belugablue}
  \verb|finp_earlytolate|} and {\small \color{belugablue}
  \verb|finp_latetoearly|}, encoding the correspondence between free
input transitions in the early semantics and input transitions in the
late semantics. The correspondence between the other types of actions
is performed analogously with two functions for each case; since their
formalization is straightforward, it is omitted.
\lstinputlisting[linerange={FinpLeft-End}]{BelugaCode/early-late-equivalence.bel}
\lstinputlisting[linerange={FinpRight-End}]{BelugaCode/early-late-equivalence.bel}
In the second statement, the input name {\small \verb|Y|} needs to be
passed as an explicit argument, otherwise Beluga would not be able to
reconstruct it during some further calls of this lemma. The proofs of
both lemmas are straightforward inductions on the given derivation.

 We can now state the signature of the functions {\small \color{belugablue} \verb|tau_earlytolate|} and {\small \color{belugablue} \verb|tau_latetoearly|}, which encode the equivalence of the two semantics:
\lstinputlisting[linerange={Thm1-End}]{BelugaCode/early-late-equivalence.bel}
\lstinputlisting[linerange={Thm2-End}]{BelugaCode/early-late-equivalence.bel}
The proof follows by induction on the given transition. In case the transition is obtained through a {\small \textsc{S-Com}} or {\small \textsc{S-Close}} rule, we apply the previously defined lemmas in order to turn an early input/output transition into a corresponding late input/output transition and then conclude.
 

\end{document}